\documentclass[sn-mathphys]{sn-jnl}
\usepackage{color}
\usepackage{cancel}
\usepackage{mathtools,cuted}
\usepackage{booktabs}
\usepackage{amsmath}
\usepackage{siunitx}
\usepackage{xr}
\newcommand\givenbase[1][]{\:#1\lvert\:}
\let\given\givenbase

\DeclarePairedDelimiterX\Basics[1](){\let\given\sgiven #1}
\jyear{2021}%
\theoremstyle{thmstyleone}%
\theoremstyle{thmstyletwo}%
\theoremstyle{thmstylethree}%
\usepackage{hyperref} 
\begin{document}

\title[Origins of Face-to-face Interaction with Kin in US Cities]{Origins of Face-to-face Interaction with Kin in US Cities}
\author[1]{\fnm{Jericho} \sur{McLeod}}\email{jmcleod3@gmu.edu}
\equalcont{These authors contributed equally to this work.}
\author[1]{\fnm{Unchitta} \sur{Kan}}\email{ukanjana@gmu.edu}
\equalcont{These authors contributed equally to this work.}
\author[1]{\fnm{Eduardo} \sur{L\'opez}}\email{elopez22@gmu.edu}

\affil[1]{\orgdiv{Department of Computational and Data Sciences}, \orgname{George Mason University}, \orgaddress{\street{4400 University Dr}, \city{Fairfax}, \postcode{22030}, \state{Virginia}, \country{United States}}}

\abstract{
People interact face-to-face on a frequent basis if (i) they live nearby and (ii) make the choice to meet. The first constitutes an \textit{availability} of social ties; the second a \textit{propensity to interact} with those ties. Despite being distinct social processes, most large-scale human interaction studies overlook these separate influences. Here, we study trends of interaction, availability, and propensity across US cities for a critical, abundant, and understudied type of social tie: extended family that live locally in separate households. We observe a systematic decline in interactions as a function of city population, which we attribute to decreased non-coresident local family availability. In contrast, interaction propensity and duration are either independent of or increase with city population. The large-scale patterns of availability and interaction propensity we discover, derived from analyzing the American Time Use Survey and Pew Social Trends Survey data, unveil previously-unknown effects on several social processes such as the effectiveness of pandemic-related social interventions, drivers affecting residential choice, and the ability of kin to provide care to family.
}


\maketitle

\section{Introduction}\label{sec:intro}
A defining feature of people's lives are the pattern of how they socialize face-to-face. These patterns are pivotal to people's economic productivity~\cite{Bettencourt}, the social support they receive for companionship or needs-based care~\cite{DunbarSpoor,Wellman,RobertsPsyc}, a sense of belonging to their local community~\cite{SCANNELL20101,LEWICKA}, and even the epidemiology of infectious diseases of their cities of residence~\cite{anderson1991infectious,VespignaniReview}. Such patterns of interaction arise from two factors, the mere presence of specific social ties living near a person regardless of interaction, and the temporal dynamics of actually engaging with those present ties for one or more activities. Here, we refer to the first of these factors as the \textit{availability} of ties, and the second as the \textit{propensity to interact} with available ties. Once availability and propensity come together, they generate observed interaction, ultimately the variable that matters for many purposes.  
For a given geographic location (e.g., a city), availability and propensity can be thought of as average per-capita quantities characterizing the location. Furthermore, it is in principle possible for friends, family, co-workers, or other types of ties to be more available on average in one location than another, as well as to have different average local propensities to engage with types of ties. If indeed such differences exist, the patterns of interaction across locations will change accordingly. These effects, as we explain in this paper, can have consequential implications. Here, to both show the detectable impacts of availability and propensity, and to argue in favor of the essential value of tracking them separately, we study the city-specific patterns of people's face-to-face interaction with extended family ties in the US. We call these ties non-coresident local family (nclf), which are family members by blood or marriage that live in the same city but not the same household. 

The relevance of cities in the study of interaction is clear: cities constitute the settings within which people conduct the overwhelming majority of their work and social lives, which also means that the consequences of face-to-face interaction are most palpable within those cities. These consequences, mentioned at the outset, encompass the economic~\cite{Bettencourt,Altonji}, social~\cite{DunbarSpoor,Wellman,RobertsPsyc,COMPTON2014,taylor1986patterns,SCANNELL20101}, and epidemiological~\cite{VespignaniReview}. The focus on extended family, on the other hand, can be explained because of their preferential status among social ties~\cite{DunbarSpoor,Wellman,personal-networks} and overwhelming abundance (at least $75\%$ of people in the US report to have nclf~\cite{Choi,Spring}). Compared to other types of social ties typically considered in surveys, nclf ties are a particularly interesting and impactful subject of study for their surprising neglect in the literature in the face of solid evidence of their critical role~\cite{Furstenberg,Bengtson} and, most urgently, the important revelation of their power to remain active~\cite{Feehan} and very risky~\cite{Cheng,Koh} during the COVID-19 pandemic.

Therefore, in order to test our theory, in this article we empirically and theoretically study the large-scale patterns of face-to-face interaction with nclf of people living in US cities. With the aid of the American Time Use Survey (ATUS) and Pew Research Center’s Social Trends Survey (PSTS) data, we develop a thorough characterization of nclf interaction, availability, and interaction propensity. We begin by showing the significant amount of nclf interaction as compared to non-family interaction, providing justification for our close attention to nclf ties. We find that nclf interaction systematically decays with city population size with virtual independence on the activity (tracked by the ATUS) driving the interaction. Similarly, we observe this decaying behavior in the availability of nclf ties as a function of city size. To discern whether the decay in nclf interaction is driven by availability or propensity, we construct a probabilistic model based on the propensity to interact given availability, allowing us to calculate how population size may affect propensity. Strikingly, we find that the propensity to interact with nclf for many types of activities is roughly constant or slightly increasing with city population size, implying that what dictates whether or not people see non-coresident family is simply that they are locally available. Consistent with this observation, we also find that the interaction duration of activities with nclf (also captured by the ATUS) is not typically affected by population size. Finally, we provide summary estimates of propensities that show which activities and times of the week (weekdays or weekends) contribute most to nclf interaction.

The interest we place in relating nclf availability and propensity to city population size and specific activities requires some elaboration. 
The first, city population, is a variable known to affect many aspects of cities including people's average productivity~\cite{Bettencourt}, the average number of social ties people have~\cite{Kasarda}, residential mobility decisions~\cite{Reia2022}, and the basic reproduction number (denoted $R_o$) of contagious diseases~\cite{Dobson}, to name a few. These effects are also known to relate to social interaction (with nclf and other ties), which means that to properly understand interaction one must relate it to population size. With respect to the activities people engage in with nclf, these are at the heart of the benefits provided by socializing with family, as stated above. Thus, for example, knowing the propensity to interact with nclf for care-related activities provides, when combined with availability, a clear picture of how much family assumes supporting roles in the lives of their kin, and how the social and economic consequences of those decisions may affect outcomes. In other words, in order to properly assess the social benefits of nclf, an empirical understanding of the patterns of interaction per activity is critical. 

At first glance, it may appear that a decomposition of interaction into a set of existing available ties and a separate propensity of encounters with those ties is an unnecessary complication, especially if what matters is their combined product, interaction. However, each of these variables functions differently. Available ties depend on people's residential location, whereas the propensity of encounters are driven by needs that are dealt with over short time scales. This means that if a need to change interaction rapidly were to arise, it would be propensity that should be adjusted. For example, the COVID-19 pandemic was marked by numerous restrictions to mobility~\cite{Hale2021} which modify propensities but not availability. Another illustration of the importance of this decomposition is highlighted by the availability of family links: if we compare two cities with different availabilities of family ties, needs for certain types of assistance such as care for young children~\cite{ogawa1996family,COMPTON2014} or the elderly~\cite{taylor1986patterns,MCCULLOCH1995} can be met by family members more often where availability is larger.

Our work offers several important contributions. Perhaps the most critical is conceptual, by introducing and supporting with evidence the idea that interaction at the level of a city can be decomposed into the more primitive components, availability and propensity. In particular, we show for nclf that indeed these two quantities can behave independently of each other and thus generate non-trivial patterns of interaction. In doing this, we offer the first systematic attempt to empirically characterize interaction and its propensity with nclf over a large collection of activities across a substantial sample of US cities, preserving both city and activity individuality. It is worth noting that, since propensity is not directly tracked in survey data, our estimation on this novel quantity generates a new angle by which to think about the practical temporal aspects of face-to-face interaction. From the perspective of each activity, we provide an overall picture of what brings nclf together with greater frequency, and we do this in a way that distinguishes such contact city by city. The implications of our findings take on several forms, inherently connected to the differences in the way that availability and propensity operate. For example, in the COVID-19 pandemic, the very existence of larger availability of nclf with decreasing population suggests that smaller cities have an intrinsic disadvantage when trying to reduce interaction given that the critical and generally favored family tie is more abundant. Another suggestion from our work is that the ability for kin to support each other varies city by city along with the benefits of this support. Finally, two other contributions that we expand upon in the \hyperref[sec:discussion]{Discussion} pertain to (i) the indication that our work makes in terms of how surveys of face-to-face interaction would improve greatly if they captured separately availability and propensity and (ii) the way in which availability and propensity relate and possibly advance other research fields such as network theory in their approach to interaction. 

\section{Results}\label{sec:results}
\subsection{Interaction}\label{sec:interaction}

We define nclf interaction, denoted $f$, as the proportion of people in the target population of the ATUS in a city $g$ that engage in an activity with nclf on a given day. Here, we consider nclf to be any family member identified by the ATUS as not residing in the same household as the respondent (see Supplementary Table~2). Because the choice to interact generally depends on the nature of the activity (e.g., care versus social) and the type of day of the week in which it occurs (weekday or weekend), we define a 2-dimensional vector $\alpha = (\text{activity, day-type})$, which we call activity-day, and assume that $f$ is a function of both $g$ and $\alpha$. We use data from the ATUS, which surveys Americans about how they spend their time and with whom during the day prior to their survey interview, to estimate nclf interaction (see \hyperref[sec:data]{Data}). Explicitly,
\begin{equation}\label{eq:fg}
    f(g,\alpha)=\frac{\sum_{i\in g}w_i a_i(\alpha)}{\sum_{i \in g} w_i},
\end{equation}
where the sum is over respondents $i$ who reside in $g$, $a_i(\alpha)=1$ if the respondent reports an $\alpha$ with nclf in the ATUS and is 0 otherwise, and $w_i$ denotes the respondent sampling weights whose unit is persons and whose sum is an estimate the target population in $g$ of the ATUS (non-institutionalized civilians aged 15 and above). The sampling weights $w_i$ can be thought of as the number of people $i$ represents in the population given our inability to survey the entire population. The weights are recalibrated so that the demographic distribution of the sample matches closely that of the target population in each city (see \hyperref[sec:reweighing]{Weight re-calibration}). Out of $384$ US metropolitan core-based statistical areas (CBSAs, or more commonly referred to as metro areas), our ATUS sample makes it possible to analyze $258$ which account for close to $260$ million people, or approximately $80\%$ of the US population over the year of the ATUS we focus on ($2016$ to $2019$, inclusive). In the ATUS, certain CBSAs are unidentified due to privacy standards. Table~\ref{tab:regres} lists the types of activities captured by the ATUS.

It is informative to begin our results with an overall estimate of the amount of interaction with nclf in the ATUS cities captured here. For this purpose, we apply Eq.~\ref{eq:fg} to those cities and find that, on average, $23.73\%$ of people in the target population interact with nclf on any given day. This can be compared to the corresponding $45.84\%$ of non-family interaction (see Supplementary Fig.~S5). While nclf interaction is estimated to be about half the interaction with that non-family contacts, one should bear in mind that non-family includes ties such as co-workers who are encountered with high frequency over the work week. Assuming that the non-family interaction estimate indicates contact on approximately a weekly basis, we see that over the cities represented here, nclf interaction occurs at a rate of about $23.73/45.84\approx 0.518$ compared to non-family. In other words, these numbers suggest that \textit{on average} people see nclf every other week.

In order to develop a more nuanced picture of nclf interaction, we now apply more detailed analyses. To understand the broad population-size effects on nclf interaction, we analyze $f(g, \alpha)$ against $p(g)=\log_{10}P(g)$, the log of population of $g$ (we use log-population due to the heavy-tailed distribution of city population in the US~\cite{Malevergne,IOANNIDES,Levy,Eeckhout2004,Eeckhout2009}).
The resulting set of points $\{p(g),f(g,\alpha)\}_g$ over all the $g$s in our data can be analyzed in several ways in order to look for possible trends. 
Here, we apply three approaches that generate complementary information: non-parametric modal regression~\cite{ChenMode}, weighted cubic smoothing splines~\cite{wang2011smoothing}, and weighted least squares. The first and second methods capture the typical and average behavior, respectively, while the last method corroborates our results.  In all cases, because the variance in the data is sensitive to sampling, we weigh each location $g$ by the sample size of the survey in that location. 

\begin{figure*}[ht!]
	\centering
		\includegraphics[width=\textwidth]{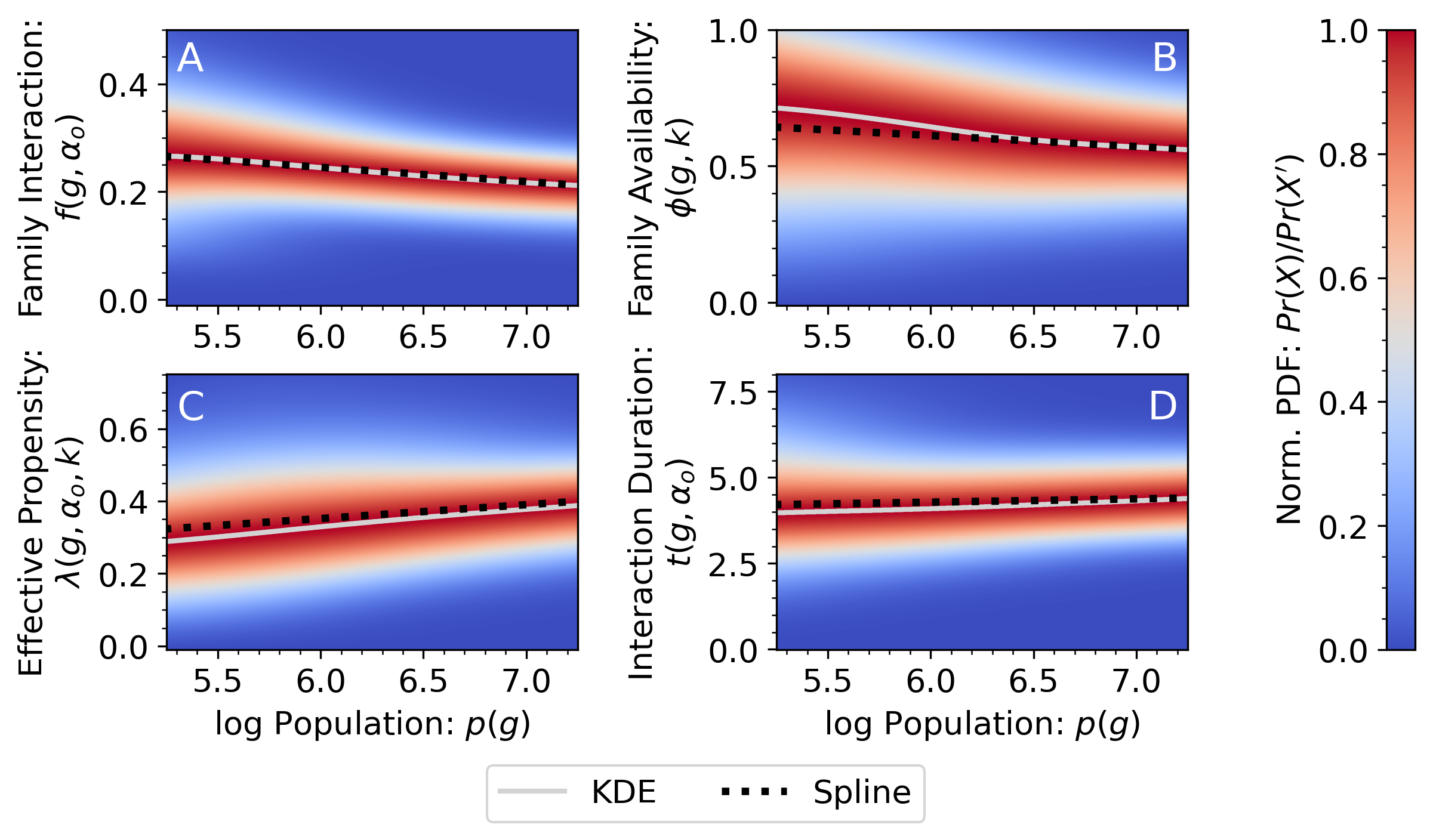}
	  \caption{Heat maps of normalized densities, conditional on city population, of interaction $f$ (panel A), availability $\phi$ (B), propensity $\lambda$ (C), and interaction duration $t$ (D). Densities for each of the quantities are calculated using pairs of points, one for each city $g$, where the horizontal coordinate is $p(g)$ and the vertical coordinate is the quantity ($f$, $\phi$, $\lambda$, and $t$) evaluated at $g$. The densities are generated using kernel density estimation (KDE). The normalization of the conditional density is done with respect to the probability of the mode of said density. Red represents large values of the normalized density (approaching $1$) whereas blue represents small values. For the relevant panels, $\alpha=\alpha_o$, which represents any activity done on any type of day (weekday or weekend) with nclf. The light gray line in each plot corresponds to the modal regression line of each plot (the line of most probable values of the normalized conditional density). The black dashed line corresponds to the smoothing spline for the same points that generate the heat map. While $f$, $\phi$, and $\lambda$ are normalized quantities ranging from $0$ to $1$, and representing a fraction of people, $t$ is a time, measured in hours. Both interaction (panel A) and availability (B) show decaying trends with population, apparent from the decay of the modal line, the smoothing spline, and the red bands. On the other hand, propensity and duration of interaction both increase with population. All relevant displayed plots are parameterized with $k=3$ in $\phi$, the number of nclf contacts in a city one considers to be needed for nclf interaction to occur. Robustness checks on the effects of survey sample size are provided in Supplementary Section~2.} 
	  \label{figs:4_panel_KDE}
\end{figure*}

We first apply non-parametric modal regression to identify the \textit{typical} behavior of $f$ with respect to $p$. This regression is based on estimating through kernel density estimation (KDE) the conditional probability density $\rho(f\given p,\alpha)$ and its conditional mode $f^*(p,\alpha)$, the value of $f$ given $p$ for which $\rho(f\given p,\alpha)$ is largest. We first apply the method to the most general $\alpha=\alpha_{o}=\text{(any activity, any day)}$; this choice of $\alpha$ also provides a useful baseline against which to compare all other results. The light-gray curve in Fig.~\ref{figs:4_panel_KDE}A shows the mode $f^*$ as a function of $p$. We can clearly observe the systematically decaying trend of $f^*(p,\alpha_o)$ as a function of $p$. For the smallest log-population $p(g)\approx 4.958$, $f^*(g,\alpha_o)$ is estimated to be $0.267$; for the large limit $p(g)\approx 7.286$, $f^*(g,\alpha_o) \approx 0.211$. This represents an overall drop of $\approx 0.056$, i.e., about $21\%$ of interaction over the log-population range. The concrete form of the trend of $f^*$ with $p$ appears to be slightly slower than linear.

Figure~\ref{figs:4_panel_KDE}A also displays a heatmap of the conditional probability density $\rho(f\given p,\alpha_o)$ scaled by $\rho(f^*\given p,\alpha_o)$. The color represents a normalized scale across values of $p$ of the location of the probability mass (note that $\rho(f\given p,\alpha_o)/\rho(f^*\given p,\alpha_o) = 1$ when $f=f^*$). The concentration of $\rho(f\given p,\alpha_o)$ above and below $f^*(p,\alpha_o)$, expressed by the intense red color, crucially suggests that $f(g,\alpha_o)$ is typically quite similar to $f^*(p(g),\alpha_o)$, the point on the modal regression 
corresponding to the population $p=p(g)$ of $g$. In addition, since $f^*(p(g),\alpha_o)$ has a systematically decaying trend with $p$, it means that in general $f$ also decays with $p$.

Aside from the typical behavior of $f$ with $p$, we also analyze its \textit{average} behavior via the well-known method of cubic smoothing splines~\cite{wang2011smoothing}. In this method, the average behavior is captured by the function $f^{(b)}(p\given\alpha_o)$ that minimizes the sum of quadratic errors between $f^{(b)}(p\given\alpha_o)$ and the data plus a cost for curvature in $f^{(b)}(p\given\alpha_o)$ which controls for over-fitting. The fitted spline is shown as the black dotted curve in Fig.~\ref{figs:4_panel_KDE}A, and exhibits a similar decay pattern as the modal regression. 

\begin{figure*}
	\centering
		\includegraphics[width=\textwidth]{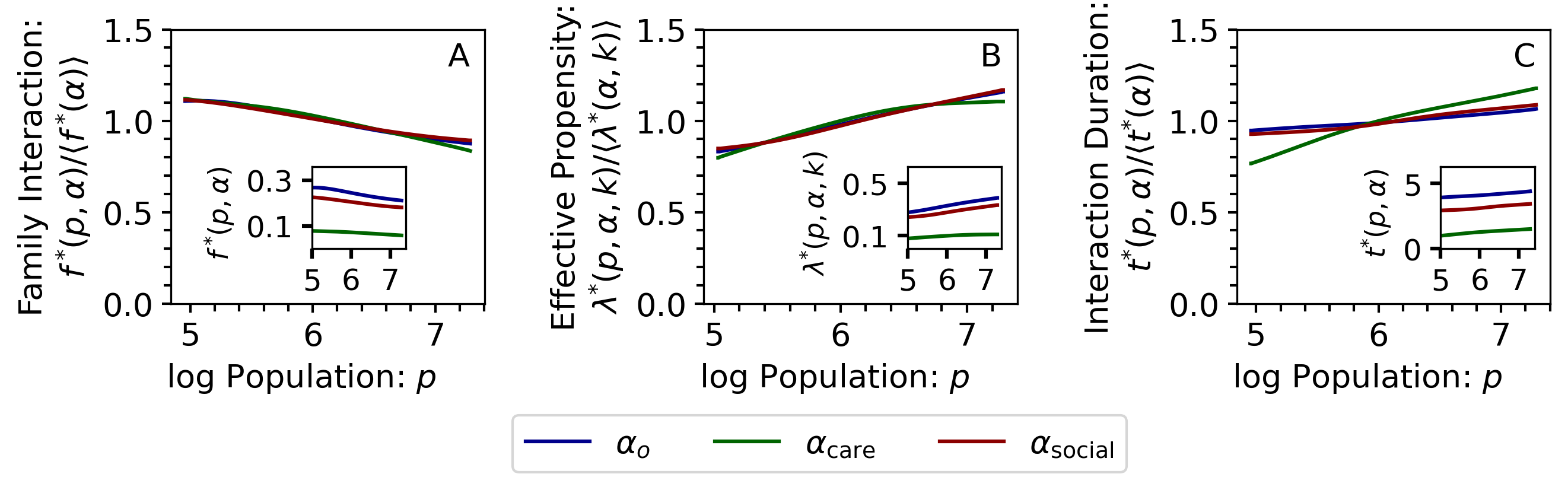}
  \caption{
  Shown in A) is the scaled KDE modal regression for family interaction, $f^*\!(p,\alpha)$, for select $\alpha$, scaled to the expected value $\langle f^*\!(\alpha)\rangle$ of each $\alpha$. The inset plot shows the corresponding result without scaling. Similarly, B) illustrates the same for effective propensity, $\lambda\!^*\!(p, \alpha, k)$, in the primary and inset plots, and C) shows shows the corresponding results for $t^*\!(p, \alpha)$ in hours. Note that $\lambda$ is parameterized with $k=3$. 
  Panels A and B show complete collapses (overlaps) of the lines in their scaled versions indicating that the population trends of both $f^*\!(p,\alpha)$ and $\lambda\!^*\!(p, \alpha, k)$ are not greatly affected by $\alpha$, suggesting an underlying common trend independent of $\alpha$. In Panel C, $\alpha$ associated with care is the only modal line that fails to collapse, which suggests that the actual duration of care activities may last longer as population increases, which may reflect additional behavioral mechanisms in need of further research.
  }
	  \label{figs:3_panel_scaled}
\end{figure*}

The decay of $f^*$ and $f^{(b)}$ with respect to $p$, observed in Fig.~\ref{figs:4_panel_KDE} for $\alpha_{o}$ is not an isolated effect.
We now expand our analysis to include `social' and `care' activities reported by the ATUS respondents, defined as $\alpha_{\text{social}}=\text{(any social activity, any day)}$ and $\alpha_{\text{care}}=\text{(any care activity, any day)}$, respectively (see Supplementary Table~4 for the ATUS activity types we consider to be social or care-related).
Using these new aggregate activity-days ($\alpha_{\text{social}}$ and $\alpha_{\text{care}}$) leads to similarly decaying modal regressions with respect to $p$. These can be seen in Fig.~\ref{figs:3_panel_scaled}A (inset), where we can also notice that, depending on the specific $\alpha$, the range of values of $f^*(p,\alpha)$ also changes.
For example, $f(g, \alpha_{\text{social}})$ ranges from about $0.226$ to $0.181$ as population increases, a range that is not very different from that of $f(g, \alpha_{o})$, whereas $f(g, \alpha_{\text{care}})$ is constrained to the range from $0.078$ to $0.058$ over the population range. 

Following the conceptual framework we proposed in the \hyperref[sec:intro]{Introduction}, we would want to know whether these decaying trends in $f$ with $p$ originate from availability or propensity. The fact that these trends persist across three different activity-days is suggestive of the hypothesis that availability is the origin of the consistent decay because it is independent of timing and activity by definition. Under this hypothesis, propensity would explain the relative differences in interaction given a specific activity-day based on people's likelihood of interacting with nclf given the nature of the engagement. If availability exhibits a decaying trend, the combination of $\alpha$-independent availability and $\alpha$-dependent propensity could generate the observed interaction behavior. To test this possibility with $f^*$ and the $\alpha$s already in Fig.~\ref{figs:3_panel_scaled}A (inset), we perform a scaling of $f^*(p,\alpha)$ by its average, given by $\langle f^*(\alpha)\rangle:=\int_{p_{\min}}^{p_{\max}} f^*(p,\alpha)dp/(p_{\max}-p_{\min})$. The rescaling leads to the main part of Fig.~\ref{figs:3_panel_scaled}A in which we can see that the curves for the three different $\alpha$ overlap (collapse). The collapse suggests that the decaying trends for each of the $\alpha$s are not unrelated but, instead, are driven by a shared function that is independent of $\alpha$ (if it was $\alpha$-dependent, the functional forms of $f^*(p,\alpha)/\langle f^*(\alpha)\rangle$ with respect to $p$ would be different and, hence, not collapse). Although this result is not in itself proof that the hypothesis just presented is correct, it does suggest that independent analyses of availability and propensity are warranted. 
The last method of analysis we employ to study the relation between $f(g, \alpha)$ and $p(g)$ is weighted least squares (WLS). First, we corroborate that the set of aggregate activities (Tab.~\ref{tab:regres}, top section), corresponding to the three cases of $\alpha$ discussed above, display negative significant regression coefficients ($\beta_1$). Note that, since $f^*(p,\alpha)$ in Figs.~\ref{figs:4_panel_KDE}A and~\ref{figs:3_panel_scaled}A shows a non-linear decaying trend, there is value in going beyond the analysis of averages based on WLS and smoothing splines. 

\begin{table}
\setlength{\tabcolsep}{5.4pt}
\begin{tabular}{llrrrlr}
\toprule
                                &  $\beta_1$         & Adj. $R^2$ &  0.025 & 0.975 &  $\beta_1$ & $\Sigma_i a_i(\alpha)$\\
\midrule
\textbf{Aggregate activity}           & $f(g,\alpha)$      &          &        &        &  \hspace{1mm}$\lambda(g,\alpha)$\\  
\midrule
\hspace{3mm}  Any                     & -0.031***          &    0.116 & -0.041 & -0.021 &    \hspace{1mm}0.032** &   7694 \\
\hspace{3mm}  Social                  & -0.026***          &    0.099 & -0.036 & -0.017 &    \hspace{1mm}0.027** &   6554 \\
\hspace{3mm}  Care                    & -0.015***          &    0.070 & -0.021 & -0.008 &      \hspace{1mm}0.002 &   2366 \\
\midrule
\textbf{Individual activity}          & $f(g,\alpha)$      &          &        &        &  \hspace{1mm}$\lambda(g,\alpha)$\\ 
\midrule
\hspace{3mm}      Social \& Leisure   & -0.024***          &    0.106 & -0.032 & -0.015 &     \hspace{1mm}0.017* &   5312 \\
\hspace{3mm}     Care, Non-Coresid    & -0.010***          &    0.043 & -0.016 & -0.005 &      \hspace{1mm}0.004 &   1849 \\
\hspace{3mm}     Eating \& Drinking   &  -0.009**          &    0.016 & -0.017 & -0.001 &   \hspace{1mm}0.026*** &   3892 \\
\hspace{3mm}            Household     & -0.009***          &    0.033 & -0.014 & -0.003 &      \hspace{1mm}0.005 &   1813 \\
\hspace{3mm}         Care, Coresid    & -0.006***          &    0.033 & -0.009 & -0.002 &     -0.003             &    623 \\
\hspace{3mm}             Traveling    &    -0.005          &    0.004 & -0.012 &  0.002 &   \hspace{1mm}0.026*** &   2951 \\
\hspace{3mm}    Consumer Purchases    &  -0.005**          &    0.015 & -0.009 & -0.000 &      \hspace{1mm}0.006 &   1134 \\
\hspace{3mm}             Religious    & -0.004***          &    0.026 & -0.007 & -0.001 &    -0.004*             &    394 \\
\hspace{3mm}    Sports, Exrc. \& Rec. &  -0.004**          &    0.014 & -0.007 & -0.000 &      \hspace{1mm}0.001 &    485 \\
\hspace{3mm}             Volunteer    &  -0.002**          &    0.016 & -0.003 & -0.000 &     -0.001             &    125 \\
\hspace{3mm}                  Work    &    -0.000          &   -0.004 & -0.002 &  0.002 &      \hspace{1mm}0.002 &    238 \\
\hspace{3mm}         Personal Care    &    -0.000          &   -0.003 & -0.001 &  0.001 &      \hspace{1mm}0.000 &     45 \\
\hspace{3mm}           Phone Calls    &    -0.000          &   -0.004 & -0.002 &  0.001 &      \hspace{1mm}0.001 &    102 \\
\hspace{3mm}    Govt Serv. \& Civic   &  \hspace{1mm}0.000 &   -0.000 & -0.000 &  0.001 &     \hspace{1mm}0.001* &     15 \\
\hspace{3mm}       Household Svc.     &  \hspace{1mm}0.000 &   -0.003 & -0.001 &  0.001 &      \hspace{1mm}0.001 &     43 \\
\hspace{3mm} Prof. \& Pers. Svc.      &  \hspace{1mm}0.000 &   -0.003 & -0.001 &  0.002 &      \hspace{1mm}0.002 &    162 \\
\hspace{3mm}             Education    &  \hspace{1mm}0.001 &    0.005 & -0.000 &  0.002 &     \hspace{1mm}0.002* &     37 \\
\midrule
\textbf{Community}                    & $\phi(g,k=3)$      &          &        &        &\\ 
\midrule
\hspace{3mm}  Family Availability     &  -0.048**          &    0.023 & -0.086 & -0.010 &      \hspace{3mm}--    &   1706 \\
\bottomrule
\end{tabular}
\caption{Weighted Least Squares analysis of the dependence of interaction $f$, availability $\phi$, and propensity $\lambda$ of cities as functions of their log-population sizes. Aggregate $\alpha$s are shown at the top, followed by specific ATUS $\alpha$, and finally, $\phi$ calculated with $k=3$ from the PSTS. For this table, all $\alpha$ distinguish activities but combine type of day of the week (weekday and weekend). Most slope coefficients associated with $f$, and all the significant ones, are negative. In contrast, the majority of coefficients for $\lambda$ are positive, with only one significant coefficient with a small negative slope (corresponding to interaction with nclf for religious activity, a trend that may reflect cultural trends). Significance levels correspond to *$p<.10$, **$p<.05$, and ***$p<.01$.}
\label{tab:regres}
\end{table}
Given the consistency uncovered for the aggregate activities, it is pertinent to explore whether decaying trends also hold for other ATUS major category activities. There are three reasons for this: (i) aggregations of several activities could be susceptible to Simpson's paradox, generating spurious trends by way of aggregation, (ii) if indeed a decaying trend in availability is affecting $f(p,\alpha)$ systematically over $\alpha$, further evidence would be garnered by doing this analysis, and (iii) for different reasons, both modal regression and smoothing spline results are not as reliable for $\alpha$s that do not occur often (see last column of Tab.~\ref{tab:regres}). The $\beta_1$ column of Tab.~\ref{tab:regres} under the area of ``Individual activity'' shows these results, which further support our thinking. Indeed, the regression coefficients for many individual activities are negative and statistically significant. The preponderance of significant negative slopes indicates that the trends observed using aggregate $\alpha$s (such as for Figs.~\ref{figs:4_panel_KDE} and~\ref{figs:3_panel_scaled}A) are not affected by aggregation effects, strengthening the case for an $\alpha$-independent availability that decays with $p$. 
Before proceeding to test availability, we note that the decaying trend in nclf interaction with population is not due to an overall decay of interaction with all non-household contacts (family or not) in big cities. On the contrary, data shows that interaction with non-household contacts in fact increase as cities get larger (see Supplementary Section~3). 

\subsection{Availability}
Although the ATUS does not provide data to directly test any trend of nclf availability with $p$, this can be done by leveraging a survey from the Pew Research Center called the Pew Social Trends Survey (PSTS) \cite{PEWsocialtrends}. As part of the PSTS, respondents are asked how many family members live within an hour's drive of where they live, providing a way to measure nclf availability. The PSTS reports this number for each respondent in discrete categories ($0$, 1 to 5, 6 to 10, 11 to 15, 16 to 20, and 21 or more).

In establishing a link between interaction and availability, a pertinent concept that should not be overlooked is that individuals do not engage with their family members on equal footing, nor do they interact with all of them~\cite{DunbarSpoor,MOK2007}—an important finding from the literature on ego networks concerned with how much interaction people have with their kin and other kinds of contacts. Instead, individuals place some kinship ties at a high level of importance while relegating others to be of low relevance. To illustrate, while a distant relative may be proximal to somebody in a location, this proximity may play no role because the relative is not particularly important in the person's social network. In contrast, interacting with, say, a parent, an offspring, a sibling, or the progeny of these family members, is likely to be much more important. This sorting, regardless of which relationships end up being important and which do not, leads to the effect that not every single kinship-tie is necessarily useful to count.

Guided by this theory, we define for each PSTS respondent $i$ a variable $b_i(k)$ that takes on the value $1$ when they report having $k$ or more nclf available, and $0$ if they report having less than $k$ nclf. Then, we define a measurement of overall \textit{availability} for each location $g$ given by
\begin{equation}\label{eq:phinf}
    \phi(g,k)= \frac{\sum_{i\in g}w_i b_i(k)}{\sum_{i\in g} w_i}.
\end{equation}
Similar to the ATUS, each PSTS respondent is given a weight $w_i$ that balances the demographic distribution of the sample such that the sample is representative at the national level. Given the categorical reporting of the PSTS variable, we develop an algorithm that allows us to reliably change $k$ by increments of one unit (see Supplementary Section~4.2). 

In order to determine the population trend, if any, of $\phi(g,k)$ as a function of $p(g)$, we carry out the modal regression and smoothing-spline analysis again. The results are shown in Fig.~\ref{figs:4_panel_KDE}B. As conjectured, the trends of availability captured by $\phi^*(p,k)$ and $\phi^{(b)}(p,k)$, respectively the modal regression and the spline of $\{p(g),\phi(g,k)\}_g$, are decaying with respect to $p$. As an additional consistency check for this decay, we calculate the WLS slope coefficient of $\phi(g,k)$ as a function of $p(g)$ and find that it is both negative and significant (see $\beta_1$ in Tab.~\ref{tab:regres}, community section). The results presented here, and for the rest of the paper, are for $k = 3$. However, varying $k$ does not change the results qualitatively (see Supplementary Figure~6).

The results here support our hypothesis that nclf interaction rates in larger cities are lower than smaller cities because overall nclf availability is also lower in the larger cities, but they do not yet paint a complete picture. One pending and interesting question is how people \textit{choose} to interact with family that is local to them (their propensity). One should contemplate the possibility that this propensity may also display negative $p$-dependence (i.e., that people in larger cities have lower need or desire to see their family). The results of Sec.~\ref{sec:interaction}, and perhaps our experience, would suggest that such a behavior may be negligible or even unlikely. Furthermore, consideration of the various functions that nclf perform runs counter to a propensity that decays with $p$. Whilst this simple picture is convincing and in agreement with intuition, none of our analysis so far can determine this. To be able to test propensity, we next introduce a probabilistic framework that can help us discern between behavioral and non-behavioral effects. 

\subsection{Probabilistic framework for interaction, propensity, and availability}\label{sec:model}

To understand the interplay between availability and propensity that leads to actual interaction, we introduce a probabilistic model where each of these effects is explicitly separated. The model is structured such that it is easy to relate to typical survey data such as what we use here. An in-depth explanation of the model can be found in Supplementary Section~5. 

Consider a model with various cities. In a city $g$, any given individual $i$ is assigned two random variables, one that indicates if the individual has nclf available ($b_i=1$) or not ($b_i=0$), and another that indicates if they report performing $\alpha$ with nclf ($a_i(\alpha)=1$) or not ($a_i(\alpha)=0$). In addition, individuals in a location $g$ are grouped into population strata within which they all share the same personal characteristics. 
For example, $i$ may be male, of a certain age range, and a given ethnicity. These characteristics are captured in the vector $\mathbf{c}(i)$. All individuals that share the same vector $\mathbf{c}$ of characteristics represent a segment of the target population, and this induces a set of weights $w(\mathbf{c})$ for each of those individuals such that the sum $\sum_{i;\mathbf{c}(i)=\mathbf{c}}w(g,\mathbf{c}(i))$ is the size of the target population $Q(g,\mathbf{c})$ in $g$ with features equal to $\mathbf{c}$. 

For simplicity, we first work with given values of $g$, $\mathbf{c}$, and $\alpha$. We introduce the probability $\kappa(g,\mathbf{c},\alpha)$ that an individual in stratum $\mathbf{c}$ of location $g$ reports doing $\alpha$ with nclf available to them. We think of this as the \textit{pure propensity} to interact. On the other hand, the probability of available nclf is given by $\phi(g,\mathbf{c})$, which does not depend on $\alpha$. While dealing with given $g$, $\mathbf{c}$, and $\alpha$, we simply use $\kappa$ and $\phi$ and then reintroduce $\mathbf{c}$, $g$, and $\alpha$ when needed. On the basis of these definitions, the joint probability that a given individual has concrete values $a_i,b_i$ is given by
\begin{equation}\label{eq:Pri}
    {\rm Pr}(a_i,b_i)=\kappa^{a_i}(1-\kappa)^{1-a_i}\phi\delta_{b_i,1}+\delta_{a_i,0}(1-\phi)\delta_{b_i,0},\quad[a_i,b_i=0,1],
\end{equation}
where $\delta_{u,v}$ is the Kronecker delta, equal to $1$ when $u=v$, and $0$ otherwise. This expression captures all possible combinations of availability and interaction: people without availability have probability $1$ to respond $a_i=0$, whereas those with availability, which occurs with probability $\phi$, have a chance of $\kappa\phi$ to report interacting with family and $1-\kappa\phi$ to report not interacting with family. 

On the basis of the personal probability captured in Eq.~\ref{eq:Pri}, and using $s=s(g,\mathbf{c})$ to represent the number of individuals surveyed in $g$ with $\mathbf{c}$, we can determine the marginal distribution that $y=y(g,\mathbf{c},\alpha)$ individuals report doing $\alpha$ with nclf, equivalent to the probability that there are exactly $y$ individuals in total for whom $a(\alpha)=1$. This marginal is given by the binomial distribution
\begin{equation}\label{eq:Pry}
    {\rm Pr}(y)=\binom{s}{y}(\kappa\phi)^y(1-\kappa\phi)^{s-y}.
\end{equation}
It is worth remembering that this applies to a specific subset of people in $g$, i.e. those respondents with personal features $\mathbf{c}$. 

We now use Eq.~\ref{eq:Pry} and the fact that its expectation is given by $y(g,\mathbf{c},\alpha)=s(g,\mathbf{c})\kappa(g,\mathbf{c},\alpha)\phi(g,\mathbf{c})$ to estimate the model expectation for actual interaction, $f_m(g,\alpha)$, which can in turn be related to our data. This calculation is straightforward because the statistics for each $\mathbf{c}$ are independent. Specifically, since each individual with $\mathbf{c}$ in $g$ represents $w(g,\mathbf{c})$ others, $f_m(g,\alpha)$ is equal to the weighted average of the expectations $y(g,\mathbf{c},\alpha)$, or

\begin{equation}\label{eq:fmg}
    f_m(g,\alpha)=\frac{\sum_{\mathbf{c}}w(g,\mathbf{c})s(g,\mathbf{c})\kappa(g,\mathbf{c},\alpha)\phi(g,\mathbf{c})}{\sum_{\mathbf{c}}w(g,\mathbf{c})s(g,\mathbf{c})}=\frac{\sum_{\mathbf{c}}Q(g,\mathbf{c})\kappa(g,\mathbf{c},\alpha)\phi(g,\mathbf{c})}{Q(g)},
\end{equation}
where the relation between sample weights, size, and population $Q(g,\mathbf{c})=w(g,\mathbf{c})s(g,\mathbf{c})$ has been used, together with the fact that $Q(g)=\sum_{\mathbf{c}}Q(g,\mathbf{c})$. Equation~\ref{eq:fg} can be interpreted as a trial of the current model, with expectation given by Eq.~\ref{eq:fmg}.

\subsection{Effective propensity}
Ideally, we would like to determine the propensity $\kappa(g,\mathbf{c},\alpha)$ for all $g$, $\mathbf{c}$, and $\alpha$. If data allowed it, we could achieve this by equating the expectation of $y$, given by $s(g,\mathbf{c})\kappa(g,\mathbf{c},\alpha)\phi(g,\mathbf{c})$, with its sample value $\sum_{i\in g} a_i(\alpha)\delta_{\mathbf{c},\mathbf{c}(i)}$ (the number of respondents that report doing $\alpha$ with nclf) and solving for $\kappa(g,\mathbf{c},\alpha)$. However, this strategy is hampered by the fact that we do not have enough information to determine $\phi(g,\mathbf{c})$ with sufficient accuracy.

To address this limitation, we employ a different strategy that is able to provide valuable information about propensity at location $g$ for each $\alpha$. To explain it, we begin by noting that availability can be written on the basis of $\phi(g,\mathbf{c})$ as
\begin{equation}\label{eq:phig}
    \phi(g)=\frac{\sum_{\mathbf{c}}Q(g,\mathbf{c})\phi(g,\mathbf{c})}{Q(g)}.
\end{equation}
Next, we introduce the quotient between interaction and availability calculated from the model. We call this quotient $\lambda(g,\alpha)$, the \textit{effective propensity} of interaction. It is given by
\begin{equation}\label{eq:lambda}
    \lambda(g,\alpha)=\frac{f_m(g,\alpha)}{\phi(g)}=\frac{\sum_{\mathbf{c}}\kappa(g,\mathbf{c},\alpha)Q(g,\mathbf{c})\phi(g,\mathbf{c})}{\sum_{\mathbf{c}}Q(g,\mathbf{c})\phi(g,\mathbf{c})}.
\end{equation}
In this expression, $\lambda(g,\alpha)$ is a weighted average of $\kappa(g,\mathbf{c},\alpha)$ over the part of the population of $g$ that does have nclf available (where the weights are $Q(g,\mathbf{c})\phi(g,\mathbf{c})$). In other words, it constitutes the effective average of propensity. 

Whilst Eq.~\ref{eq:lambda} allows us to interpret the meaning of $\lambda(g,\alpha)$, its calculation is straightforwardly done by using the sample values $f(g,\alpha)$ and $\phi(g)$ given respectively by Eqs.~\ref{eq:fg} and \ref{eq:phinf}. Fig.~\ref{figs:4_panel_KDE}C, we show the modal regression $\lambda^*(g,\alpha_o)$ and smoothing-spline $\lambda^{(b)}(g,\alpha_o)$ for the set of points $\{(p(g),\lambda(g,\alpha_o))\}_g$. Both curves show that population size has a very small increasing effect on $\lambda(g,\alpha_o)$, from which we can deduce that the propensity to interact with nclf for those people that have nclf available is not being limited or reduced by $p$. 

As with $f^*(p,\alpha)$, $\lambda^*(p,\alpha)$ for the aggregate $\alpha_o$, $\alpha_{\text{social}}$, and $\alpha_{\text{care}}$ also collapse when divided by their averages $\langle \lambda^*(\alpha)\rangle:=\int_{p_{\min}}^{p_{\max}}\lambda^*(p,\alpha)dp/(p_{\max}-p_{\min})$ (Fig.~\ref{figs:3_panel_scaled}B). Although this may superficially appear unsurprising given the relations between $f$, $\phi$, and $\lambda$ (Eq.~\ref{eq:lambda}), it is worth keeping in mind that $\lambda^*(p,\alpha)$ is the directly estimated conditional mode of the $\lambda(g,\alpha)$ data. Thus, the collapse suggests that the trends that govern the values of $\lambda(g,\alpha)$ for individual cities are indeed multiplicative: a product of a $\phi$ dependent on $g$ but independent of $\alpha$, and a pure propensity $\kappa$ dependent on $\alpha$,  predominantly. 

A more comprehensive analysis of trends of $\lambda$ with respect to $p$ is performed through WLS, shown in Tab.~\ref{tab:regres} (fifth column) for the remaining $\alpha$s in the ATUS. There are $8$ significant coefficients, of which only religious activities has a negative slope, albeit with a very small value; the remaining significant coefficients are positive. Among all coefficients (significant and non-significant), only $3$ coefficients altogether have negative slopes. This shows the predominant tendency for the effective propensity to either increase slightly with $p$ or be roughly independent of it.

\begin{figure}
    \centering
    \includegraphics[scale=0.8]{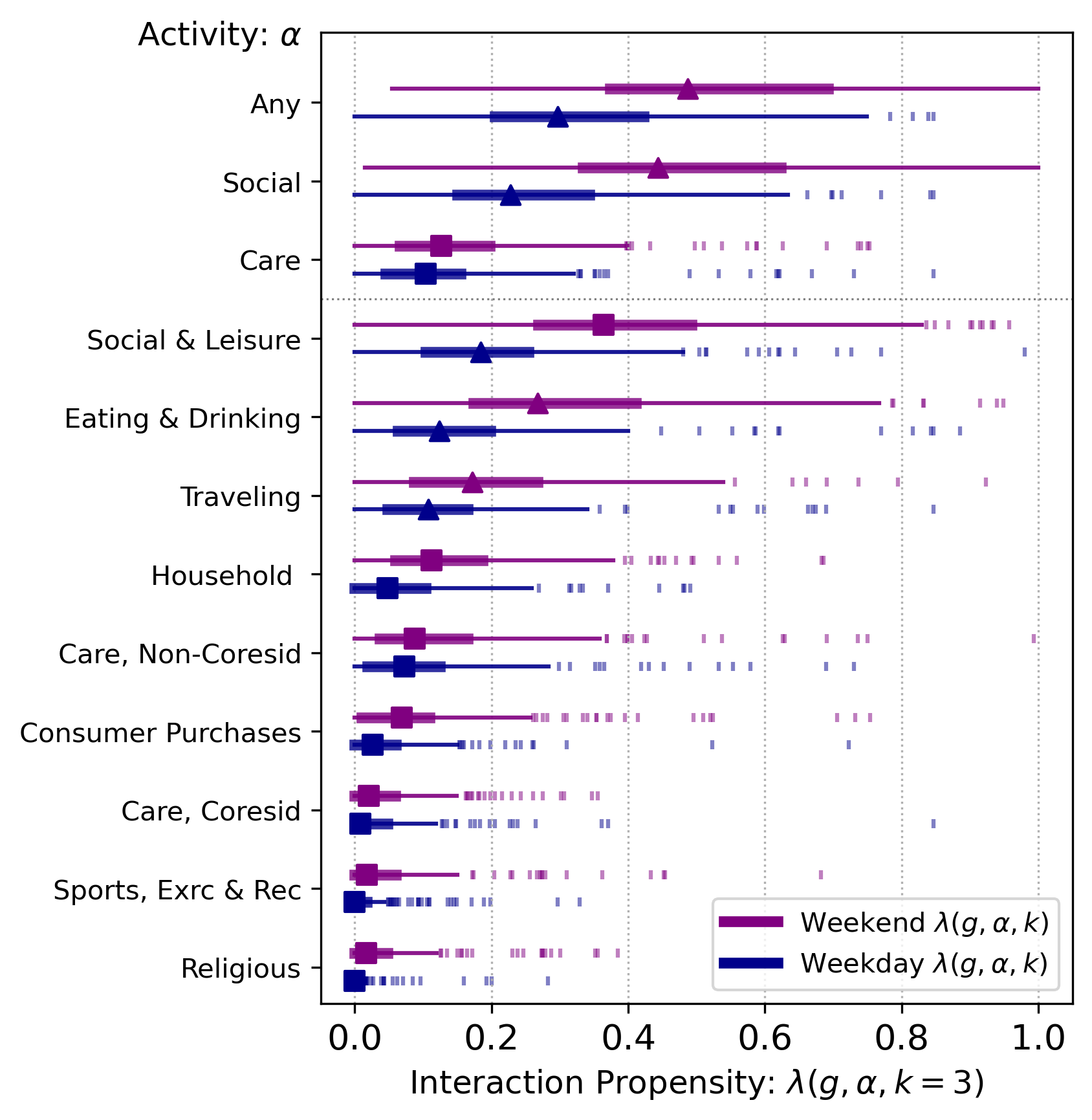}
    \caption{
    Box plots of interaction propensities $\lambda(g,\alpha,k)$ for well-sampled $\alpha$s in ATUS, along with aggregate $\alpha$ (at the top). Here, weekends and weekdays have been kept separate to distinguish the changes in propensity with respect to this variable. The triangles and squares are located at the point of median of the $\lambda(g,\alpha,k)$ for a given $\alpha$. Triangles pointing up represent an increasing trend of the propensities with $p$ and squares represent no significant increasing or decreasing trend, consistent with the results on Tab.~\ref{tab:regres}; triangles pointing down would appear if a significant decaying trend was present, but none are obtained. The bold lines represent the IQR, thin lines the minimum of the remaining range observed or $1.5\times$IQR in each direction, and the remaining points (the outliers) are denoted with pipe-marks. The selected activities were filtered to remove those with medians not noticeably above 0. These box plots show which activities respondents choose to do with family, ordered by aggregation followed by propensity.
    }
    \label{fig:lambda-boxplots}
\end{figure}

Having studied the propensity $\lambda$ with respect to $p$, we now focus on how various activity-days $\alpha$ affect it. In Fig.~\ref{fig:lambda-boxplots}, we present box plots of the sets of values $\{\lambda(g,\alpha)\}_g$ for those activity-days that are sufficiently well-sampled. Here, we make a distinction between weekday and weekend when constructing $\alpha$s. Median values are represented by the triangle and square shapes. Triangles point up if $\lambda$ of a given $\alpha$ shows a statistically significant increase with $p$ and squares indicate that $\lambda$ has no significant trend with $p$. No statistically significant decreasing trends of $\lambda$ with $p$ were observed for any $\alpha$. The trends and their significance correspond to those in Tab.~\ref{tab:regres}. Aggregate versions of $\alpha$ are placed at the top of the figure to provide a reference. As expected, people exhibit a larger propensity to interact with nclf on weekends than on weekdays for all activity categories. 
Spending social time with family has the largest propensity with weekend median value of $0.44$, which can be interpreted as the chances that an individual with nclf available would do a social activity with them on a weekend are approximately $44\%$, in marked contrast to weekdays when the propensity drops to $22.7\%$. In comparison, care-related activities have weekday propensities of $10.4\%$, increasing slightly to $12.6\%$ on the weekends. Another observation from Fig.~\ref{fig:lambda-boxplots} is that the ranges of values of $\lambda$ can be large for the most common activities. The values of $\lambda$ can be used for estimation of interaction in a variety of cases. For example, assuming independent random draws with $\lambda$ as the success rate for any given $\alpha$, we can quickly estimate such quantities as the proportion of people (in a city or nationally) that meet with nclf in a period of time (say, a month) to do $\alpha$, or the average wait time until the first meeting to do $\alpha$.

To complete our analysis of $\lambda$, we present two other summary results in Supplementary Section~7. First, we provide a rank-ordered reference of activities by propensity averaged over the US in Supplementary Table~4. This shows which activities are associated with high or low propensity. Second, to learn about how values of $\lambda(g,\alpha)$ are distributed across concrete US metropolitan areas, and particularly which places exhibit either considerably larger or smaller propensities than other cities of similar populations, we also present Supplementary Table~5, which shows the top and bottom $10$ locations by average rank-order based on weighted $z$-scores of $\lambda(g,\alpha)$ with respect to $\lambda^*(p(g),\alpha)$, where ranks are averaged for values of k ranging from 1 to 3. Similar analysis is conducted for interaction duration $t$ (defined next) and shown in Supplementary Table~6.

\subsection{Interaction duration with nclf family}
As a final analysis of the interplay between people's nclf interaction and population size, we study one last quantity captured in ATUS: the \textit{interaction duration} with nclf. The relationship between interaction duration and population size can also be studied with the techniques we have used thus far. 

Let us denote the duration of $i$'s interaction with nclf for $\alpha$ by $t_i(\alpha)$. Thus, for a given location $g$ and $\alpha$, the average interaction duration is given by
\begin{equation}
    t(g,\alpha)=\frac{\sum_{i\in g}w_i t_i(\alpha) a_i(\alpha)}{\sum_{i\in g}w_i a_i(\alpha)},
\end{equation}
where $a_i(\alpha)$ is defined the same way as in Eq.~\ref{eq:fg}. Note that interaction duration is averaged only over those that report nclf interaction for the $\alpha$ under consideration. 
This is because $t$ clearly involves a behavioral component, and we are interested in determining the role that population size may exert in the behavior of people when they interact with nclf. If an increase in $p$ was associated with a decrease in the duration of interaction, it could suggest, for example, that people's busy lives in larger cities limit how much they are able to interact with family, which could be a signal associated with the decaying trends for $f$. 

In Figs.~\ref{figs:4_panel_KDE}D and~\ref{figs:3_panel_scaled}C we present examples of modal regressions and smoothing splines of interaction duration $t$ with respect to $p$ for aggregate $\alpha_{o}$, $\alpha_{\text{social}}$, and $\alpha_{\text{care}}$. In all cases, $t^*(g,\alpha)$ and $t^{(b)}(g,\alpha)$ are either approximately steady over $p$ or increasing slightly. As in the case of propensity, interaction duration captures the behavioral dimension of interaction. The lack of a decaying trend provides further support for the notion that family availability may be the main driver of interaction decay and that people's attitudes towards interacting with family are not diminished by population size.

\section{Discussion}\label{sec:discussion}

The separation of interaction into two necessary factors, availability and propensity, provides a new lens by which to understand it in a more coherent and principled way. Over relatively short periods of time (say, weeks or months) and at the population level of cities, patterns of availability are rigid, which is to say they are structural and approximately fixed in time and space. On the other hand, patterns of propensity are due to day-to-day decision-making and encompass the bulk of the agency that individuals have in controlling interaction at any given time in the short term; propensity is much more flexible than availability, and generally operates at shorter time scales. If we phrased interaction in the language of social networks, availability can be thought of as a static well-defined social network based purely on underlying formal social relations (unambiguously defined for nclf, e.g., parent, sibling, in-law), while propensity is a stochastic process occurring on the static network. 

Perhaps the most salient observation we make is that, at the level of population of cities, availability explains the heterogeneity in nclf interaction, effectively providing clear support for our approach. In addition, for most activities propensity shows either a weak positive population-dependent trend or no trend. This supports another one of our most important conclusions that, when family is locally available, people take ample advantage of their presence to interact with them. Moreover, the time invested in those interactions are not negatively impacted by population size (i.e., the busy lives of residents of big cities do not deter them from spending time with family). Phrased in social network theory terms, our results imply that the static social networks of nclf differ by city and those differences are the main drivers of the differences in face-to-face interaction we observe, and that nclf propensity is roughly independent of the structure of these networks. 

This robust family-interaction effect may have important consequences in how people shape the rest of their social networks~\cite{DAVIDBARRETT201720,personal-networks}. At a more fundamental level, if availability plays a dominant role in nclf interaction, the patterns that characterize this availability are also likely to characterize other critical aspects of life that strongly depend on nclf interaction (say, care-related outcomes~\cite{COMPTON2014,taylor1986patterns,MCCULLOCH1995} or certain aspects of people's well-being~\cite{DunbarSpoor,MOK2007,roberts2015managing}). 

As suggested in the \hyperref[sec:intro]{Introduction}, another aspect that depends on interaction is propagation of infectious diseases. Current research in epidemiological modeling does not pay attention to the concepts of availability and propensity when dealing with interaction data. Consider, for example, the large scale European POLYMOD survey concerned with determining population mixing matrices capturing patterns of social contacts~\cite{Mossong}, frequently used in population-level epidemiological modeling in Europe. Availability is not tracked in this survey (or most other population-level surveys of interaction), but such information can prove critical when large disruptions to regular patterns of behavior (i.e. propensity) can take place, such as has been observed during the COVID-19 pandemic. For one, availability in most cases would not change in the rapid way propensity can change. Also, the way that propensity can adjust after the start of a pandemic is constrained by availability; i.e., availability is a template inside of which propensity can adapt to pandemic conditions. While not all study of human interaction is aimed at such extremes, the framework of availability and propensity still applies. Furthermore, because the fundamental concepts of availability and propensity are not restricted merely to family ties, our remarks here are relevant to social interactions generally.

All the examples provided above can lead to policy considerations. Many non-pharmaceutical interventions applied during the COVID-19 pandemic have been directed at reducing propensity for face-to-face interaction or aiding those with economic needs (see~\cite{PERRA20211} for a review of intervention), but they have hardly considered availability, which may reflect structural needs of the populations of those places. In low-income areas, family support to take care of young children~\cite{COMPTON2014} is an economic necessity. Asking people in such locations to stop seeing family may prove ineffective. In places with low nclf availability, the closure of daycare during the COVID-19 pandemic coupled with the lack of family support could lead to more mothers dropping out of the workforce, perpetuating gender inequality and slowing down economic recovery. Thus, policy interventions that consider these needs at the local level would likely have a better chance at being effective~\cite{Hale2021}. From the standpoint of other support roles performed by nclf, a better measurement and understanding of the interplay between availability and interaction may inform how to best address the needs of people at a local level in ways not unlike those that address housing, education, or other locally-oriented policy making. 

The origin of the population pattern displayed by availability, although not our focus here, is likely related to domestic migration~\cite{litwak1960occupational,litwak1960geographic}. The more pronounced absence of family in larger populations may be driven by the fact that larger cities tend to offer a variety of opportunities for work, education, and access to particular services and amenities not always available in cities with smaller populations, thus generating an in-migration that is, by definition, a source of decreased family availability. In this scenario, one of the main drivers of the population trend in availability would be economic (see similar examples~\cite{KANCS2011191}), suggesting a feedback mechanism whereby the in-migration leading to less family availability can be improving or sustaining economic success, in turn perpetuating the incentives for that in-migration over time. However, other mechanisms such as a desire to maintain close ties with family are likely to balance the economic incentives, preventing the feedback from becoming a runaway effect~\cite{LEWICKA,Spring,litwak1960geographic,litwak1960occupational}. Note that this last point does suggest a coupling between availability and propensity that operates at \textit{long} time scales (say, over one or multiple years), by which the fraction of the population that is not willing to be far from extended family would exercise agency to adjust their living conditions through migration to be closer to family. 

Our study contains certain limitations. First, while demographic characteristics are incorporated into our model, estimates of propensity and availability for specific demographic strata are not obtained here due to limitations in survey sampling. Such demographic understanding of interaction is important and should be considered in future research if appropriate data can be obtained. Second, although the ATUS tracks a substantial variety of activities, one should not view it as a comprehensive catalog of all possible ways in which people may interact. To name one example, activities that involve more than a single day are not currently collected in the ATUS by design. Other such limitations exist and researchers should be mindful of that. Third, it may prove fruitful to consider \textit{material} availability that could affect interaction (for example, in a city without a cinema an $\alpha$ for going to the movies would have propensity $=0$). We do not expect this factor to play a major role for the $\alpha$s and locations we study, as our smallest cities have populations of $\approx 130,000$ people. However, investigations focused on more detailed activities in smaller places could require taking this effect into account. Fortunately, the structure of our model can be straightforwardly updated to include such material availability given sufficient data.

At a methodological level, there are interesting observations our work suggests. For example, some simple modifications to large-scale surveys of people's interaction and time-use behavior could lead to extremely beneficial information able to enhance the usefulness of such surveys. The ATUS could add a few simple questions to their questionnaires that would determine whether respondents have family, friends, or other contacts locally available to them \textit{even if no activities have been performed with them}. Such questions could provide baseline information about not just what people do but their preferential choices, helping to distinguish the effects of both propensity and availability on interaction. 

As a final reflection, we relate the notions of availability and propensity to the very successful discipline of network science and its application to spreading processes on social networks~\cite{newman2018networks,VespignaniReview,Havlin,HOLME201297,porter2016dynamical}. In this theoretical context, our concept of availability translates to a static network structure of social connections that exist independent of contact activity. Propensity, on the other hand, would be represented by a temporal process of contact activity occurring on the static network. In this context, what the results of the present work mean for the study of processes on networks is that realistic models need to take into account the heterogeneity in both the types of ties (network links) and their associated contact processes. In the case of epidemics, the introduction of frameworks such as those in~\cite{Karrer,Karsai} offers great promise because they are compatible with the non-trivial structure we uncover here. This contrasts with most other literature that basically conceives of what we call propensity as a low-probability, permanent contact process occurring with all contacts of the static network (the Markovian assumption~\cite{VespignaniReview}), a modeling choice that destroys most of the true complexity of the process. Dedicated data that addresses propensity will go a long way in driving the development of a new generation of more realistic models of propagation processes on networks.

In summary, we decompose face-to-face interaction with non-coresident local family at the city level in terms of availability and propensity to interact, and find that while availability decays with the population of those cities, neither the propensity to interact across activities nor the duration of those interactions shows the same decay. The decay in availability is sufficient to lead to an overall decay of interaction with nclf across US cities as their population increases. We arrive at these results by introducing a stochastic model that allows us to combine existing survey data, the American Time Use Survey and Pew Research Center's Social Trends Survey, to estimate availability and propensity at the US metro level. Analysis of the resulting propensities show that social activities are the most common, especially on weekends. In social networks terms, availability can be thought of as static social networks of family relations and propensity as a process on these networks. Our findings indicate that the availability network differs by city and is the main driver of the variance in observed face-to-face interaction, while propensity is roughly independent of the network structure. We also discuss some of the implications of our framework and offer ideas on how it is relevant for survey-design, scientific research, and policy considerations with some particular attention to the context of the COVID-19 pandemic.

\section{Material and methods}\label{sec:matmethods}
\subsection{Weighted least squares}\label{sec:WLS-and-Weights}
Since location sample sizes diminish with population, much of our analysis utilizes weightings. One method we employ is weighted least squares (WLS), defined on the basis of the model
\begin{equation}
    v_g=\beta_1 u_g +\beta_o + \epsilon_g,
\end{equation} 
where $g$ indexes the data points, and the error terms $\epsilon_g$ do not necessarily have equal variance across $g$ (heteroskedastic). The solution to the model is given by the values of the coefficients $\beta_1$ and $\beta_o$ such that $\sum_g w_g\epsilon_g^2$ over the data points is minimized. The $w_g$ correspond to $g$-specific weighing used to adjust how much the regression balances the importance of the data points.

The Gauss-Markov Theorem guarantees that the weighted mean error $\sum_g w_g\epsilon_g^2$ is minimized if $w_g=1/\text{var}(\epsilon_g)$. On the other hand, it is known that the variance in a survey is inversely proportional to the size of the sample. Therefore, in our analysis we use
\begin{equation}\label{eq:wg-samp}
    w_g=\frac{1}{\text{var}(v_g)}=\frac{s(g)}{\sum_g s(g)}\quad(\text{for }v_g=f(g,\alpha),\phi(g),\lambda(g,\alpha),t(g,\alpha)),
\end{equation}
where $s(g)$ is the number of respondents in $g$. These weights apply to $f(g,\alpha)$ from the ATUS as well as $\phi(g)$ from the PSTS, where the same $g$ may have different relative weights based on the samples collected for each survey in the same location $g$.

To compute these weights when the dependent variable is $\lambda(g,\alpha)$, we make use of the technique called propagation of error. For $\lambda(g,\alpha)=f(g,\alpha)/\phi(g)$, its variance is given as a function of the variances of $f(g,\alpha)$ and $\phi(g)$, calculated through a Taylor expansion up to order $1$. In the limit of weak or no correlation between $f(g,\alpha)$ and $\phi(g)$ (which is reasonable here given that the variables are independently collected),
\begin{equation}
    \text{var}(\lambda(g,\alpha))\approx \left(\frac{\partial \lambda}{\partial f}\right)^2\text{var}(f(g,\alpha))+\left(\frac{\partial \lambda}{\partial\phi}\right)^2\text{var}(\phi(g)).
\end{equation}
Assuming the covariance between $f(g,\alpha)$ and $phi(g)$ is negligible, we calculate
\begin{equation}
    w_g\approx\left[\left(\frac{\partial \lambda}{\partial f}\right)^2\text{var}(f(g,\alpha))+\left(\frac{\partial \lambda}{\partial\phi}\right)^2\text{var}(\phi(g))\right]^{-1}\quad(\text{for }\lambda(g,\alpha)),
\end{equation}
where both $\text{var}(f(g,\alpha))$ and $\text{var}(\phi(g))$ are given by the Eq.~\ref{eq:wg-samp}. 

\subsection{Weighted cubic smoothing splines}
One of the techniques we employ to approximate the functional dependence between $p$ and, separately, $f$, $\phi$, $\lambda$, and $t$ is cubic smoothing splines~\cite{wang2011smoothing}, defined as follows. For two generic variables $u_g$ and $v_g$, the cubic smoothing spline $v^{(b)}(u_g)$ is a piecewise smooth function that minimizes 
\begin{equation}
    \sum_g w_g(v_g-v^{(b)}(u_g))^2+\eta\int_{\min_g\{u_g\}}^{\max_g\{u_g\}} \left(\frac{d^2v^{(b)}}{d\xi^2}\right)^2d\xi,
\end{equation}
where $\eta$ is a penalty parameter that controls how much to discourages $v^{(b)}$ from large amounts of curvature (and thus overfitting), and $\xi$ is a dummy variable. In the limit of $\eta=0$, curvature is not penalized and the algorithm overfits the data; when $\eta\to\infty$, the only possible solution requires $d^2 v^{(b)}/d\xi^2\to 0$, leading to a straight line and thus making the algorithm equivalent to weighted least squares (WLS). In the algorithm we employ~\cite{WOLTRING1986104}, the function $v^{(b)}$ is constructed from cubic polynomials between the data points along $\{u_g\}_g$, with the condition that they are smooth up to the second derivative along consecutive pieces.

The penalty parameter $\eta$ can be either chosen arbitrarily or, instead, determined on the basis of some selection criterion. In our case, we use generalized cross-validation~\cite{Golub}, which optimizes prediction performance. The algorithm is implemented as the function \texttt{make\_smoothing\_spline} of the package \texttt{scipy.interpolate} in \texttt{python}~\cite{splines}.

\subsection{Modal regression using kernel density estimation}
Another method we employ to estimate how population $p$ affects $f$, $\phi$, $\lambda$, and $t$ is non-parametric modal regression.~\cite{ChenMode}. Intuitively, the method looks for the \textit{typical} behavior of a random variable as a function of some independent variable.

This method is defined in the following way. Using a smoothing kernel (in our case Gaussian), we construct the $2$-dimensional kernel density estimator $\rho(u,v)$ for the set of data points $\{(u_g,v_g)\}_g$~\cite{hastie2009elements}, where
we choose bandwidth by inspection in the neighborhood of the Silverman method~\cite{silverman1986density} but favor solutions that tend towards the smoothing spline results as they are a sign of a stable modal regression line (details can be found in Supplementary Section~6).
Then, we determine the conditional density $\rho(v\given u)=\rho(u,v)/\rho(u)$ and extract its local mode $v=v^*(u)$. Here, we use this method for unimodal $\rho(v\given u)$ (multimodality can be handled by the method, but is not relevant here). 

\subsection{Data}\label{sec:data}
Our interaction data is derived from the American Time Use Survey (ATUS) conducted by the Bureau of Labor Statistics (BLS)~\cite{ATUS}. Each year, the ATUS interviews a US nationally-representative sample regarding their full sequence of activities through the day prior to the interview, termed a diary day. The information collected in this process is recorded in several files, including the ``Respondent'', ``Activity'', and ``Who'' data files. We link these files for the period between 2016-2019 to get comprehensive information about each respondent, activities they carried out on the diary day, as well as those who accompanied the respondent during each activity. We consider nclf to be companions of the respondent who are family but do not reside in the same household (see Supplementary Table~2). These may include the respondent's parents, parents-in-law, own children under 18 years of age, and other family members as long as they do not reside with the respondent. Activities are encoded in the ATUS with codes of $6$ digits, the first two representing activity categories such as 'eating \& drinking', 'personal care', or 'work', to name a few. Additional digits provide more specificity about the activity such as the context or some other detail. We restrict our analysis to activities at the two-digit level (called major categories in the ATUS lexicon) which encompass seventeen such codes. The ATUS also captures the day of the week when the respondent has been interviewed, which in known to play an important role in the choices of activities people perform. We encode the combination of activity and type of day with the 2-dimensional vector variable $\alpha$. Therefore, as an example, eating and drinking with family done on a weekend day corresponds to a specific value of $\alpha$. Beyond the use of the ATUS 2-digit codes, we also create aggregate activity categories that serve as baselines to our analysis and capture people's major social functions. At the most aggregate level, we define $\alpha_{o}$ which combines all activities in the ATUS done on either type of day (weekday or weekend), which we refer to as 'any activity, any day'. We also define $\alpha_{\text{social}}$, the aggregate set of social activities with the 2-digit codes $11-14$ done any day of the week. Finally, we define $\alpha_{\text{care}}$, we combine the 2-digit codes $3-4$ on any day of the week.

To estimate local family availability, we use the Pew Social Trends Survey (PSTS) from the Pew Research Center which was conducted in 2008 using a nationally-representative sample of 2,260 adults living in the continental US~\cite{PEWsocialtrends}. The PSTS identifies geographic location of the respondents at the county level and the binned quantity of family members who live within a one-hour driving distance. To work with the data at the CBSA level, we map county FIPS codes to CBSA codes using the crosswalk downloadable from the National Bureau of Economic Research~\cite{nbercrosswalk}. After filtering by the set of CBSAs common to both surveys, our final sample size is $N=$ 30,061 for the ATUS and $N=$ 1,706 for the PSTS.

Metro populations are obtained from the American Community Survey (ACS) data (5-year estimates, 2015-2019) published by the US Census Bureau~\cite{censusacs}. We also extract various demographic variables from the US Census population estimates to use in the recalibrations of sampling weights (see \hyperref[sec:reweighing]{Weight re-calibration}).

\subsection{Weight re-calibration}\label{sec:reweighing}
Our analysis of the time variables of the ATUS is performed with consideration to sampling weights. The ATUS provides a sampling weight for each respondent which, in essence, gives an estimated measure of how many people within the US population the respondent represents given the respondent's demographic and other characteristics. (The unit of these weights is technically persons-day, although for the purpose of our study we normalize by the number of days in the survey period since it simplifies interpretation of the weights to just population, and because our sampling period is fixed.) The use of such weights is meant to reduce bias and improve population-level estimates of quantities captured by the raw survey data. The ATUS respondent weights are calibrated by BLS at the national level which can be reasonable for large cities but are not reliable for smaller CBSAs. For this reason, we perform a re-calibration of these weights at the CBSA level, the main unit of analysis in our study. 

Our methodology for weights recalibration follows the original BLS procedure which is a 3-stage raking procedure (also known as iterative proportional fitting and is widely used in population geography and survey statistics \cite{Deming1940, idel2016review}) but with constraints imposed at the CBSA level and without non-response adjustments. The goal of this type of adjustments is to find a joint distribution of weights given dimensions of characteristics (e.g., race by sex by education by age) such that the sum of respondent weights along a given characteristic axis (i.e., the marginals) matches a known control or target population. For each CBSA, the target in each stage of the 3-stage procedure corresponds to the CBSA population by sex by race by ATUS reference day, population by sex by education level by ATUS reference day; and population by sex by age by ATUS reference day, respectively. For the stratification of these characteristics, see Supplementary Table~3.

We use estimates from the US Census Bureau's Population Estimates Program as well as the ACS. Unavoidably, there are small differences in the target population used by BLS and one that is used by us (for example, population estimates from the PEP include those institutionalized and non-civilian, whereas the universe for the ATUS does not include this sub-population).

\bmhead{Supplementary information}

The present article is accompanied by a Supplementary Information document.

\bmhead{Acknowledgments}
We acknowledge helpful suggestions from Eben Kenah, Robin Dunbar, and Serguei Saavedra. No external funding source has been used in this research.

\section*{Funding}
Not applicable

\section*{Conflict of interest}
The authors declare no conflicts of interest.

\section*{Ethics approval}
Not applicable

\section*{Availability of data and materials}
All data sources are publicly available.

\section*{Code availability}
Code used in the project is mostly open source software from the \texttt{python} language projects \texttt{scipy}, \texttt{statsmodels}, and \texttt{pandas}.

\section*{Authors' contributions}
Supressed for double-blind review.

\bigskip
\begin{appendices}
\end{appendices}
\bibliography{main}
\end{document}


\flushbottom
\maketitle
%
%
\thispagestyle{empty}

\section{Assessment of ATUS data quality}

\subsection{Respondents per capita}
\begin{figure}[h]
    \centering
    \includegraphics[width=0.5\textwidth]{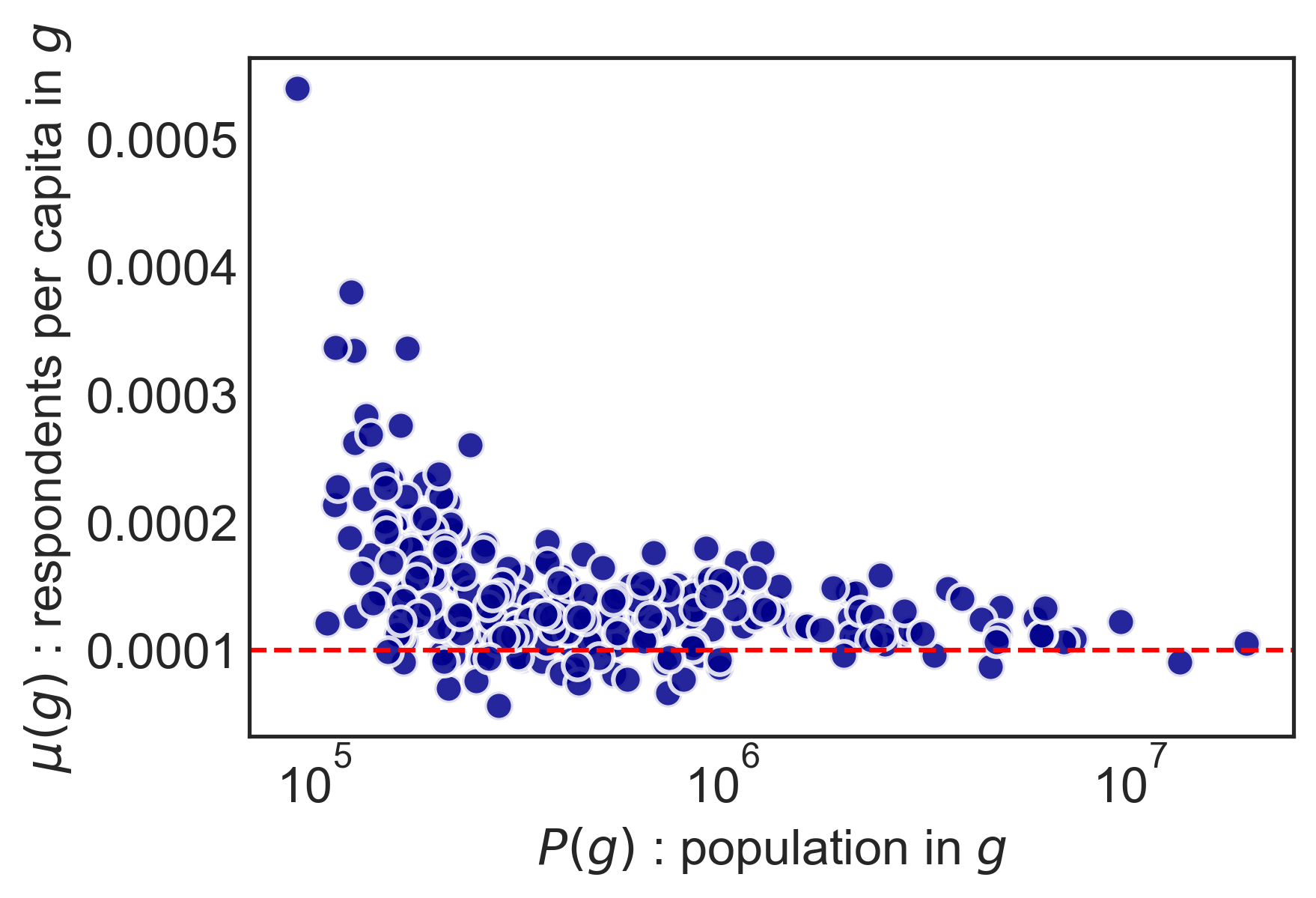}
    \caption{Repondents per capita in the ATUS for CBSAs used in our study. Each dot corresponds to a CBSA and represents its population and ATUS respondents per capita. The horizontal scale for population $P(g)$ is logarithmic in order to compensate for the long-tailed distribution of CBSA sizes in the US. It is clear that the representation rate hovers just above $10^{-4}$ (red dashed horizontal line), although it is larger for less-populated CBSAs.}
    \label{fig:mu_vs_p}
\end{figure}

To better understand the level of coverage of the American Time Use Survey (ATUS) over the CBSAs we analyze, we present a scatter plot of the number of respondents in each CBSA $g$ who are also part of our analysis (see Fig.~\ref{fig:mu_vs_p}). The plot has logarithmic horizontal scale. Each dot corresponds to a city $g$. The horizontal coordinate corresponds to population $P(g)$ and the vertical to respondents per capita $\mu(g)$, given by the ratio between the number of respondents in $g$ and the population $P(g)$.

The plot shows how the respondents per capita approach a rate typically just above $10^{-4}$ as $P(g)$ increases, that is, a $1$ in $10,000$ uniform probability over a CBSA population, with a negligible percentage of CBSAs that drop below this rate. When $P(g)$ goes below $\approx 3\times 10^5$ (i.e., for smaller CBSAs), $\mu(g)$ increases drastically. 

Note that given Fig.~\ref{fig:mu_vs_p}, it is clear that the number of respondents grows approximately linearly with the CBSA population for locations with $P(g)\gtrsim 3\times 10^5$, satisfying the equation
\begin{equation}
    {\rm respondents \ in\ }g\approx 10^{-4} \times P(g).
\end{equation}
Crucially, this relation shows that the response coverage is not biased in a way that would likely generate \textit{systematically} poorer statistics for smaller cities. 

\subsection{Non-response bias}\label{sec:nonresponse-bias}
\begin{figure}[ht]
    \centering
    \includegraphics[width=\textwidth]{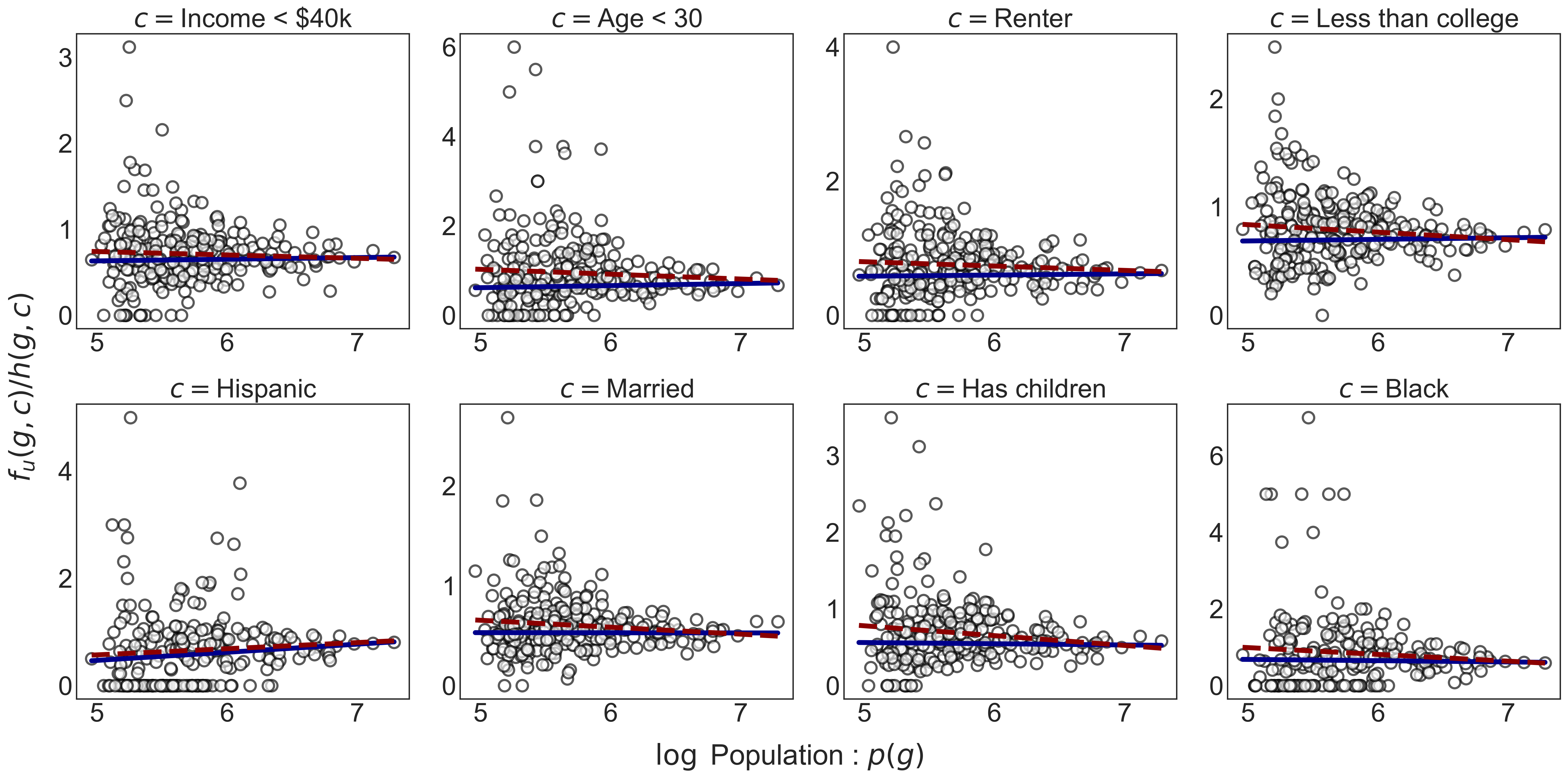}
    \caption{Scatter plot of $f_u(g,c)/h(g,c)$ and log-population $p(g)$ as well as the corresponding WLS (blue) and OLS (red) regression lines for the collection of cities we study in the ATUS. The ratio $f_u(g,c)/h(g,c)$ captures the unweighted nclf interaction rate among ATUS respondents in $g$ with characteristic $c$ with respect to the response rate of sampling units in $g$. Each panel corresponds to a different characteristic $c$, indicated in the panel title. Slope values and their statistical significance are reported in Tab.~\ref{tab:nonresponse}.}
    \label{fig:f/h}
\end{figure}

\begin{table}[ht]
\centering
\begin{tabular}{rll}
\toprule
                $c$ &     $\beta_{1,WLS}$ &  $\beta_{1,OLS}$ \\
\midrule
   Income less than \$40,000 &  \hspace{1.2mm}0.02  & -0.04 \\
            Age less than 30 &  \hspace{1.2mm}0.05  & -0.11 \\
                      Renter &  \hspace{1.2mm}0.02  & -0.06 \\
           Less than college &  \hspace{1.2mm}0.01  & -0.07 \\
                    Hispanic &  \hspace{1.2mm}0.15* &  \hspace{1.2mm}0.11 \\
                       Black & -0.03                & -0.18 \\
                     Married & -0.00                & -0.07* \\
                Has children & -0.02                & -0.13* \\
\bottomrule
\end{tabular}
\caption{Values of the slopes from WLS and OLS analyses done for Fig.~\ref{fig:f/h}. * indicates significance levels of $p<.10$.}
\label{tab:nonresponse}
\end{table}

Non-response is always a major concern of any survey if the characteristics or behaviors of non-respondents systemically differ from those of respondents, as this may introduce bias with respect to the variable of interest. For the ATUS in particular, it has been found that while there are differences in the characteristics of respondents and non-respondents, estimates of time use patterns with and without non-response adjustments are broadly comparable~\cite{Abraham}. In practice, no tests will be able to rule out the possibility of bias in a variable that is unobservable for the non-respondents. 

For the sample we use in our study, the response rate is $\approx 40.35\%$ (i.e., the non-response rate is $\approx 59.65\%$). In this section, and considering the limitations we have just pointed out, we assess whether the potential non-response bias in ATUS could be strong enough that the observed decaying trend with population in nclf interaction $f$ does not qualitatively reflect the actual trend in nclf interaction. We perform two different analyses for this assessment. 

For our first analysis, let us provide some intuition~\cite{Lohr}. Ideally, we would like to be able to see how the response propensity correlates with having an interaction with nclf among sampling units and how this correlation is related to population size. However, because we have no way of observing the interaction variable for the non-respondents, we instead look at other variables that may be associated with interaction and how respondents behave given these other variables. For example, suppose that among the respondents who have less than a college degree, the average probability that they interact with nclf decreases with population size. If the response rate among sampling units (everyone that has been sampled by the ATUS whether they responded to the survey or not) with less than a college degree also increases with population size, then it suggests that the actual trend in $f$ could either be similar to what is observed or decay more strongly, considering that non-college graduates in smaller cities are more likely to interact with nclf than non-college graduates in larger cities but they also tend to respond to the survey at a lower rate. In other words, if the non-college graduate non-respondents in these smaller cities \textit{were} to respond, $f$ may be higher for the corresponding population range. Other such scenarios are possible that would yield similar a situation.

To analyze this trend generally and for various characteristic variables, let us denote $f_u(g,c)$ as the unweighted average nclf interaction rate among respondents in $g$ who have a characteristic $c$. Let us also denote $h(g,c)$ as the response rate among the sampling units in $g$ who have a characteristic $c$. (Here we are using unbolded $c$ to indicate that we are working with a single characteristic, unlike in our interaction model in the main text and in Sec.~\ref{sec:SImodel}.) Our quantity of interest is $f_u(g,c)/h(g,c)$, which we plot against the log-population of $g$, denoted $p(g)$ as in the main text. In the example situation of non-college graduates above, or when $f_u(g,c)$ decreases with $p(g)$ faster than $h(g,c)$, we would expect $f_u(g,c)/h(g,c)$ to also decrease with $p(g)$. Such a negative relationship could suggest that the potential bias due to non-respondents with the corresponding characteristics may strengthen the observed decaying trend in $f(p)$ reported in the main text. On the other hand, if there is not a relationship between $f_u(g,c)/h(g,c)$ and $p(g)$, it may be reasonable that the non-response bias does not affect our results in a systematic way.

We first perform WLS with $f_u(g,c)/h(g,c)$ as the dependent variable and $p(g)$ as the independent (see Methods in the main text for details on WLS). We consider the following characteristics $c$: having income less than \$40,000; being 30 years old or younger, being a renter, having less than a college degree, being Hispanic, being married, having children, and being Black. As can be seen in Fig.~\ref{fig:f/h} (blue lines) as well as Tab.~\ref{tab:nonresponse}, all of the WLS coefficients are very weak (close to 0) and all but one are statistically insignificant. The only coefficient that is statistically significant is positive and corresponds to the characteristic of being Hispanic. We also repeat this analysis using OLS (Fig.~\ref{fig:f/h}, red lines) to not obscure the effect of the weights of $g$ and find that most of the coefficients are now weakly negative, although most of them remain statistically insignificant. The results from both WLS and OLS suggest that there is no systematic linear relationship between $f_u(g,c)/h(g,c)$ and $p(g)$, but if there is one, the effects are likely to be too small to qualitatively change the $f(p)$ trend observed in the main text. 

To corroborate the results from this first analysis, we perform an additional check. The ATUS is conducted via phone calls and it may be necessary for interviewers to call a respondent more than once to initiate or complete an interview. As suggested by Ref.~\cite{Abraham}, the response of having interacted with nclf or not that is obtained from respondents who required many calls can be suggestive of the behavior of non-respondents if we assume that these difficult-to-contact respondents are more similar to non-respondents. Furthermore, a high number of calls in a location may be a sign of difficulty of contact and thus undersampling in that location. As the data on the phone calls made to the respondents are published by the ATUS, we are well-situated to combine these two ideas into one statistic to diagnose the potential direction of non-response bias.

Let us define by $e_i$ the number of calls placed to respondent $i$. In a given location $g$, if the ATUS was able to sample every person, one would estimate that the total number of calls placed would be $\sum_{i\in g} w_i e_i$. Now, the number of calls we would estimate lead to an answer of `yes' to interaction with nclf is equal to $\sum_{i\in g}w_i e_i a_i$ because interaction is captured when $a_i=1$. This means that the proportion of calls \textit{needed} to discover nclf interaction in $g$ is given by $\psi(g)= \sum_{i\in g}w_i e_i a_i / \sum_{i\in g}w_i e_i.$ Conversely, $1-\psi(g)=1 - (\sum_{i\in g}w_i e_i a_i / \sum_{i\in g}w_i e_i)$ gives the proportion of calls \textit{needed} to discovering that no interaction with nclf has taken place.
Here, we use $\alpha = \alpha_o = \text{(any activity, any day)}$. 

\begin{figure}
    \centering
    \includegraphics[scale=1]{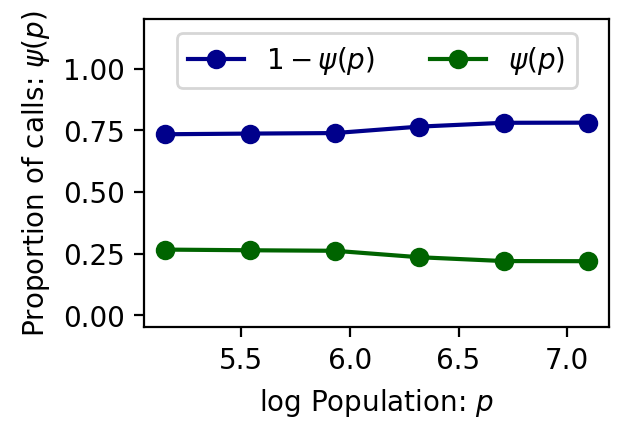}
    \caption{Proportion of calls $\psi$ over all calls placed by interviewers needed to find a respondent reporting to have had nclf contact (green curve), as a function of population. Also, the additive complement with respect to $1$ of $\psi$, i.e., the proportion of calls $1-\psi$ over all calls placed by interviewers needed to find a respondent reporting no nclf interaction (blue curve). This means that as population grows, it is proportionally easier to find respondents that report nclf interaction and harder to find respondents that report no nclf interaction. 
    }
    \label{fig:psi}
\end{figure}

In Fig.~\ref{fig:psi}, we bin $\psi(g)$ by population and show that, as CBSA population increases, it is easier to find nclf interaction and hard to find non-interaction with nclf. Following the logic in Ref.~\cite{Abraham}, this suggests that the decaying trend of $f(p)$ with respect to $p$ is likely to be even more pronounced than what we find in the current study, as interviewers have a harder time obtaining that answer as $p$ increases. In other words, this analysis suggests that our results may be a conservative estimate for the population decaying trend we observe in $f(p)$.

\section{ATUS and PSTS sample-size robustness checks}
\begin{figure}[h]
    \centering
    \includegraphics[width=0.8\textwidth]{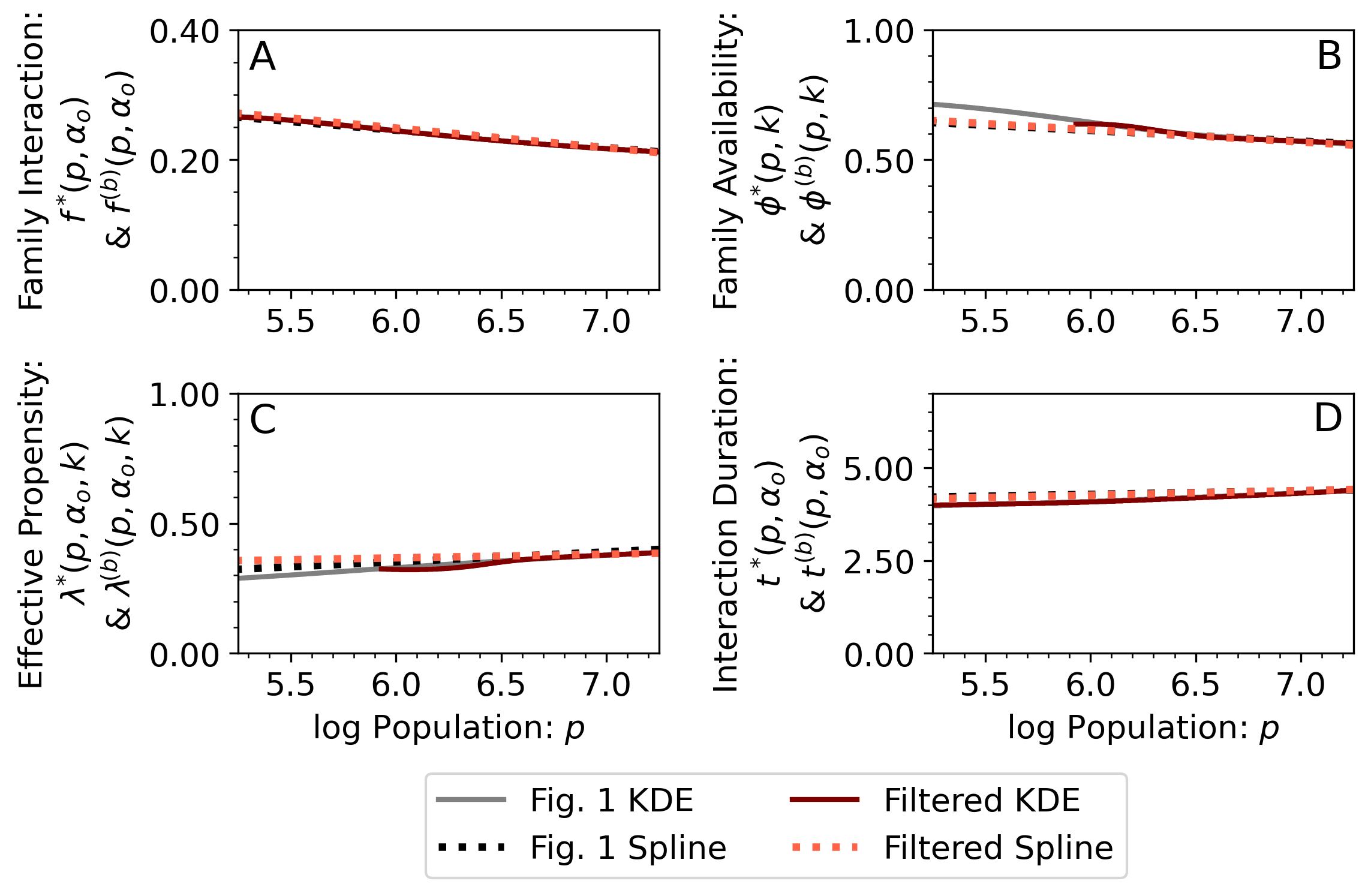}
    \caption{
    Modal regressions from Fig.~1 of the main text with the same modal regressions performed after applying filtering on respondent sampling rates. A CBSA must have at least 30 respondents in ATUS and 10 respondents in PSTS. This demonstrates the consistency of our reported results with those obtained by considering only well-sampled cities.}
    \label{fig:sampling_panels.png}
\end{figure}

To provide further confidence in the robustness of our results, we check whether the sample size of each city has an effect in the functional dependence with respect to $p$ for the modal regressions $f^*$, $\phi^*$, $\lambda^*$, and $t^*$ or smoothing splines $f^{(b)}$, $\phi^{(b)}$, $\lambda^{(b)}$, and $t^{(b)}$. In other words, we check if Fig.~1 of the main text changes if the cities chosen to create Fig.~1 are restricted to those where the sample sizes of people interviewed in each of the cities cannot be below a certain threshold. In Fig.~\ref{fig:sampling_panels.png}, we present a comparison between the modal regressions and smoothing splines in Fig.~1 of main text with the same curves made only with CBSAs where at least $30$ people are sampled in the ATUS and $10$ in the Pew Social Trends Survey (PSTS). The results are remarkably consistent. The differences in sample sizes required for the ATUS and the PSTS stem from the differences in survey sizes, which makes setting more ambitious lower bounds difficult as they lead to much smaller numbers of cities with which to analyze the data.

\section{Overall social interaction}
\begin{figure}[ht]
    \centering
    \includegraphics[scale=0.95]{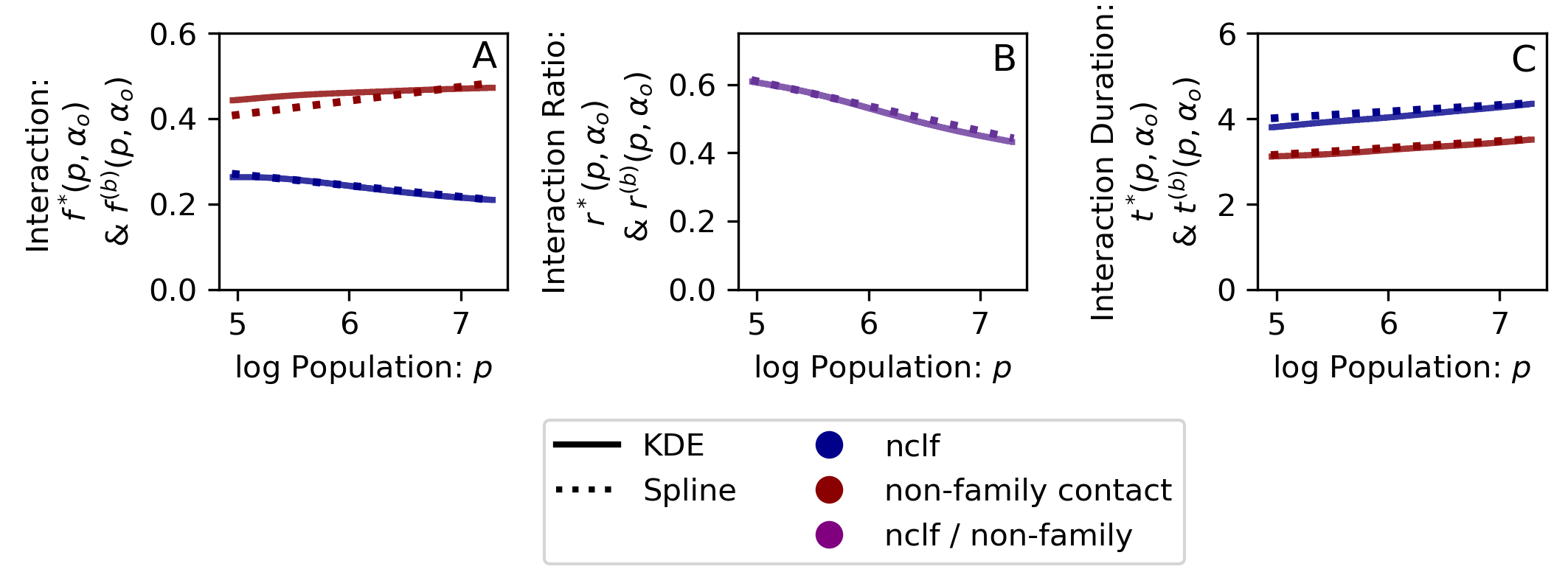}
    \caption{Here we update $a_i(\alpha)$ to be 1 if respondent $i$ has contact with the examined type of relationship; either nclf or non-coresident non-family (non-fam), and 0 if not, and otherwise hold to our prior definitions. Using the modal regressions $f^*$, $r^*$, and $t^*$, and the splines $f^{(b)}$, $r^{(b)}$, and $t^{(b)}$, to compare non-coresident family with non-coresident non-family contacts, it is apparent the trends in interactions themselves differ relative to population, while the trends in interaction duration, shown in hours, do not, precluding the ideas that people are less social or prefer to spend less time with family in cities with larger populations.}
    \label{fig:fam_nfam}
\end{figure}
To provide another robustness check on our finding of nclf interaction decay as a function of population, we measure overall social interaction with non-household contacts (not just non-coresident family) as a function of log-population. We focus only on non-work social interactions. We calculate interaction $f_{\rm non-fam}(g, \alpha_o)$ for local contacts who are both non-coresident and non-family using the same method as with nclf (Eq.~1 in the main text). The quotient between $f(g, \alpha_o)$ and $f_{\rm non-fam}(g, \alpha_o)$ is then defined by
\begin{equation}
    r(g,\alpha_o)= \frac{f(g,\alpha_o)}{f_{\rm non-fam}(g,\alpha_o)}.
\end{equation}
We study the typical behavior of $r(g, \alpha_o)$ using cubic smoothing splines and modal regression to obtain $r^{(b)}(p, \alpha_o)$ and $r^*(p, \alpha_o)$. In Fig.~\ref{fig:fam_nfam}, we show the results and see that both interaction rate (panel A) and interaction duration (panel C) with non-coresident non-family contacts (non-fam) in general increase with population. This suggests that the decaying trend in interaction of nclf with population is not due to people in larger cities being less social generally. We also see that, relative to interactions with non-family contacts (Fig.~\ref{fig:fam_nfam}B), nclf interaction markedly decreases as population increases.

\section{Defining family}\label{sec:si-availability}
\subsection{Pew Social Trends Survey Question}
Respondents from the Pew Research Center survey were asked how many extended family members live within a one-hour drive of the respondent, and their responses were binned to ranges of $0$, $1-5$, $6-10$, $11-15$, $16-20$, and $21$ or more. Even though this is not identical to asking about nclf, we consider the two meanings of extended family: family members who are outside of the nuclear family of a focal person, and family members who do not co-reside with the focal person. The second interpretation coincides with our definition with nclf. The survey questions included clarification on extended family, which defined it to include members of nuclear families, implying the intent of referring to nclf as extended family. We also examined subsequent surveys from the Pew Research Center, and note that in later surveys referring to extended family, the additional phrase "who do not live with you" is added, which highlights a transition from an implicit definition to an explicit one. 

\subsection{Estimation of family availability}\label{si:ds-n_f}

\begin{figure*}[tb]
	\centering
		\includegraphics[width=0.7\textwidth]{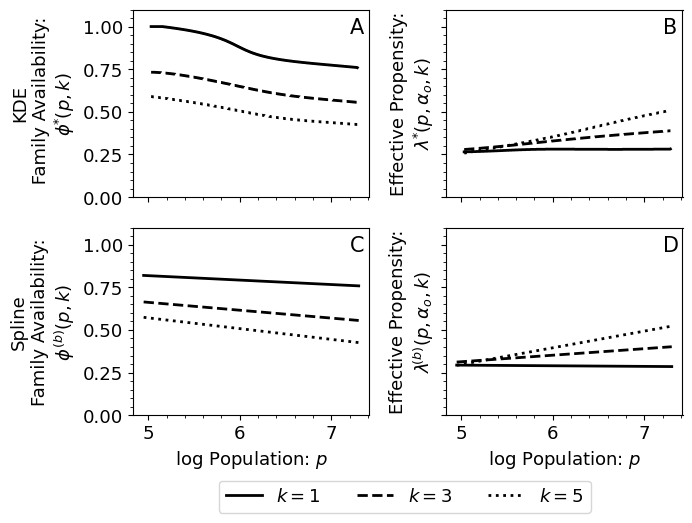}
	  \caption{The effects of varying $k$ on family availability $\phi(p,k)$ and propensity to interact $\lambda(p,\alpha_o,k)$ are shown here via $\phi^*$ and $\lambda^*$ modal regressions and $\phi^{(b)}$ and $\lambda^{(b)}$ splines. Increasing $k$ shifts down the estimates for $\phi$ but does not change the qualitative behavior of $\phi$, as shown in panels A and C. $\lambda$ depends on both $\phi$ and $f$; varying $k$ (and thus effecting $\phi$) can change the qualitative behavior of $\lambda$ as shown in panels B and D. However, regardless of which $k$ we choose, the results in all cases support our finding in the main text that it is availability that drives nclf interaction down as city population increases, not propensity.}
	  \label{figs:4_panel_n_f}
\end{figure*}

Guided by the theory of kinship in ego networks, in Sec. 2.2 of main text we define for each respondent $i$ of the PSTS $b_i(k)$ which takes on a value of 1 if $i$ reports having at least $k$ nclf available within an hour's drive and 0 otherwise. Given the categorical reporting of this survey question ($0$, $1-5$, $6-10$, $11-15$, $16-20$, and $21$ or more), we develop an algorithm in order to increment the resolution of $k$ to single units ($1,2,3,\dots$) as follows. 

Because this binning is somewhat arbitrary, we would like to obtain a reasonable estimate of the within-bin probability that a respondent has a specific quantity of nclf within an hour's drive given the data that is available. We achieve this by finding a discrete probability distribution that would also conform to the overall distribution of the categorical responses of the survey respondents given appropriate binning. Concretely, we first determine the proportion of respondents belonging to each answer category (bin). Then, we fit a negative binomial distribution such that the error between the probability mass function summed over the values in each bin (1 to 5 for the $1-5$ bin, etc.) and the proportions of respondents per bin is minimized. After the overall distribution is fitted, we derive a separate probability distribution (the negative binomial estimated in the previous step) for each bin by normalizing the probability masses of the values falling in the bin so that they sum up to 1 (note that this restricts the support of the distribution corresponding to each bin to just the values falling inside the bin.) We use these distributions to estimate the probabilities of each respondent having at least $k$ nclf members within an hour's drive for any integer $k$ value within the category reported by the respondent. 

Application of this method leads to the plots in Fig.~\ref{figs:4_panel_n_f}. Here, we also show the effects of $k$ in the estimation of $\phi(p, k)$ as well as $\lambda(p, \alpha, k)$ in Fig.~\ref{figs:4_panel_n_f}. Note that as $k$ increases, this captures the idea that, for some individuals, not all nclf reported available may be relevant to them in terms of interaction (for some people, some family members are not that important emotionally, consistent with Ref.~\cite{DunbarSpoor}). It is interesting to see, however, that even $k=1$ leads to a $\lambda^*(p,\alpha_o,k=1)$ and $\lambda^{(b)}(p,\alpha_o,k=1)$ that are non-decaying.

\subsection{Classification of family and non-family in the ATUS}

\begin{table}[t]
    \centering
    \begin{tabular}{|l|l|l|}\hline
         \textbf{ATUS Code} & \textbf{Description} & \textbf{Relationship} \\\hline\hline
         40 & Own Non-household Child & Family  \\\hline
         51 & Parents             & Family  \\\hline
         52 & Other nonhousehold family members < 18 & Family  \\\hline
         53 & Other nonhousehold family members 18 and older & Family  \\\hline
         54 & Friends             & Non-Family  \\\hline
         56 & Neighbors/acquaintances & Non-Family \\\hline
         57 & Other nonhousehold children < 18 & Non-Family  \\\hline
         58 & Other nonhousehold adults 18 and older & Non-Family \\\hline
         59 & Boss or manager     & Non-Family  \\\hline
         60 & People whom I supervise & Non-Family  \\\hline
         61 & Co-workers          & Non-Family  \\\hline
         62 & Customers           & Non-Family  \\\hline
    \end{tabular}
    \caption{Classification of ATUS relationship types~\cite{ATUSdict} into non-coresident family and non-coresident non-family in our study.}
    \label{tab:fam_def}
\end{table}

The ATUS provides context for each activity reported by a respondent by including information on who was present, the relationship type between the respondent and the companion, and whether the companion co-resides with the respondent. 
We provide a summary of the non-coresident relationship types provided by the ATUS in Tab.~\ref{tab:fam_def}, which is a direct reflection of the data dictionary provided in Ref.~\cite{ATUSdict}, and the classification we use in our analysis to identify non-coresident family and non-family. Note that a corresponding table is not required for the PSTS due to the design of the survey separating relationship types explicitly. 

For the purposes of Fig.~\ref{fig:fam_nfam}, in cases where a single $\alpha$ event occurs with both nclf and non-coresident non-family present with the respondent, we consider it for each relationship type. The presence of additional nclf or non-coresident non-family does not further affect interaction or increase the duration of a single event.

\section{Probabilistic framework for interaction, propensity, and availability}\label{sec:SImodel}

In this section, we expand on the probabilistic model introduced in the main text. The model represents a survey which is structured as the ATUS and PSTS used here, but it is assumed that respondents are being asked all the relevant questions in a single hypothetical survey. The aim of the model is to separate the effects of availability and propensity on the observed interaction, but with enough generality that from it one can derive various quantities relating back to the ATUS and PSTS.

\subsection{General Model Description}\label{sec:model}
Consider a hypothetical survey where the information gathered in both the ATUS and the PSTS was collected. We model the target population of the survey in a location $g$ as stratified into $q$ demographic strata. Each stratum is described by a $d$-dimensional vector of personal characteristics $\mathbf{c}$. As an example, one could have a sex-by-education stratum which is described by a 2-dimension vector $\mathbf{c} = (\text{sex, education})$, where sex and education can take the values in the sets \{male, female\} and \{less than high school, high school diploma or above\}, respectively (see Tab.~\ref{tab:c} for the specific characteristics used in this article connected to the ATUS, specifically relevant for the recalibration of respondent weights, main text, Methods).
The characteristics of a respondent $i$ of the survey are denoted by $\mathbf{c}(i)$.
We label $s(g,\mathbf{c})$ the number of respondents in location $g$ with specific features $\mathbf{c}$, and their respective survey sampling weights as $w(g,\mathbf{c})$. The weights are created in such a way that 
\begin{equation}\label{eq:P_ws}
    Q(g,\mathbf{c})=w(g,\mathbf{c})s(g,\mathbf{c})
\end{equation}
is satisfied, where $Q(g,\mathbf{c})$ is the size of the target population with characteristics $\mathbf{c}$ in $g$. In our model, we assume that any pair of respondents $i$ and $j$ in the same $g$ with $\mathbf{c}(i)=\mathbf{c}(j)$ have the same weight $w(g,\mathbf{c})$ (this holds for the re-calibrated ATUS weights we constructed and used in this study.)

\begin{table}[]
    \centering
    \begin{tabular}{|l|l|}\hline
         \textbf{Characteristic}&\textbf{Possible Values}\\\hline\hline
         Age& <23, 23-59, 60+ \\\hline
         Gender& male, female \\\hline
         Race/Ethnicity& non-Hispanic \& non-black alone, non-Hispanic \& black alone, Hispanic and any race\\\hline
         Education& completed high school, did not complete high school\\\hline
    \end{tabular}
    \caption{Personal characteristics for stratification of population used the ATUS weight re-calibration (see the main text). In addition, characteristics such as these are the model for the stratification into various $\mathbf{c}$ in the model discussed in Sec.~\ref{sec:SImodel}.}
    \label{tab:c}
\end{table}

The modeled survey asks respondents two questions about nclf: 1) whether they have non-coresident family available locally and 2) whether they interacted with nclf on the day prior to the survey by performing an activity-day $\alpha$ with their nclf (here, activity-day $\alpha$ is defined in the same manner as in the main text).
We now introduce for each respondent $i$ two indicator variables $a_i$ and $b_i$ that can only take two values, $0$ and $1$. Variable $b_i$ takes the value of $1$ when $i$ indicates that they have nclf available in their city, and $0$ otherwise. Similarly, $a_i=1$ when a respondent indicates they performed an activity-day $\alpha$ with nclf, and $a_i = 0$ otherwise. Both $a_i$ and $b_i$ are treated as random Bernoulli variables. Note that $b_i=0$ implies $a_i=0$ because, by construction, a respondent cannot perform an activity with nclf if there are none available. 

\subsection{Fixed city $g$, population stratum $\mathbf{c}$, and activity-day $\alpha$}
We now define the conditional probability of $a_i=1$ given that $b_i=1$ as $\kappa$ (which is implicitly a function of $g$, $\mathbf{c}$, and $\alpha$). Symbolically,
\begin{equation}
    {\rm Pr}(a_i=1|b_i=1)=\kappa.
\end{equation}
Under the same conditional assumption ($b_i=1$), the probability that the respondent $i$ replies `no' to interacting with family is $1-\kappa$, or
\begin{equation}
    {\rm Pr}(a_i=0|b_i=1)=1-\kappa.
\end{equation}
These two different situations can be written under the single expression
\begin{equation}
    {\rm Pr}(a_i|b_i=1)=\kappa^{a_i}(1-\kappa)^{1-a_i}.
\end{equation}
Conceptually, $\kappa$ is the same as the propensity to interact, a concept we discuss in the main text.

On the other hand, conditional on not having family available ($b_i=0$), the probability to respond `yes' to the nclf interaction question is $0$, and to respond `no' is $1$, which leads to
\begin{equation}\label{eq:pacondt0}
    {\rm Pr}(a_i|b_i=0)=\delta_{a_i,0},
\end{equation}
where $\delta_{u,v}$ is the Kronecker delta, equal to $1$ if the integers $u$ and $v$ are equal and $0$ otherwise. 

We also define the probability for $i$ to have nclf as $\phi$ (again dependent on $g$ and $\mathbf{c}$ but independent on $\alpha$ by definition), which is the probability that $b_i=1$. Consequently, $b_i=0$ has probability $1-\phi$. This definition allows us to write the joint probability of responding any combination of $a_i$ and $b_i$ as
\begin{equation}\label{eq:pri}
    {\rm Pr}(a_i,b_i)=\kappa^{a_i}(1-\kappa)^{1-a_i}\phi\delta_{b_i,1}+\delta_{a_i,0}(1-\phi)\delta_{b_i,0}.
\end{equation}
This equation corresponds to Eq.~3 of the main text. The marginals will be useful later, and thus, we have:
\begin{equation}
    {\rm Pr}(a_i)=\sum_{b_i=0}^1{\rm Pr}(a_i,b_i)=\delta_{a_i,0}(1-\phi)+\kappa^{a_i}(1-\kappa)^{1-a_i}\phi
\end{equation}
and
\begin{equation}
    {\rm Pr}(b_i)=\sum_{a_i=0}^1{\rm Pr}(a_i,b_i)=\phi\delta_{b_i,1}+(1-\phi)\delta_{b_i,0}.
\end{equation}

For a given set of $g$, $\mathbf{c}$, and $\alpha$, individual respondent probabilities can be combined to produce the full probability to obtain a specific set of survey replies $\{a_i,b_i\}_i\to\mathbf{a},\mathbf{b}$, which takes the form
\begin{equation}\label{eq;Prat}
    {\rm Pr}(\mathbf{a},\mathbf{b})=\prod_{i=1}^{s}\left[\kappa^{a_i}(1-\kappa)^{1-a_i}\phi\delta_{b_i,1}+\delta_{a_i,0}(1-\phi)\delta_{b_i,0}\right],
\end{equation}
where we have written $s(g,\mathbf{c})$ as $s$ for simplicity and where we have adopted the convention that respondents within $g$ and $\mathbf{c}$ can be sequentially labelled from $1$ to $s$.

One of our key goals is to estimate the unknown value of $\kappa$ through the responses in the survey, since this would let us express nclf interaction as a product of propensity and availability, thus explicitly showing their effects. To do this, we determine the probability that a total of $y=\sum_{i=1}^s a_i$ respondents say they have interacted with nclf, formally given by
\begin{equation}\label{eq:pryv1}
    {\rm Pr}(y)=\sum_{\mathbf{a}}\sum_{\mathbf{b}}\delta_{y,\sum_i^s a_i}{\rm Pr}(\mathbf{a},\mathbf{b}),
\end{equation}
where the sums over $\mathbf{a}=(a_1,\dots,a_s)$ and $\mathbf{b}=(b_1,\dots,b_s)$ correspond to all the possible configurations the responses among the $s$ respondents can take. Also, because respondent's $i$ responses are independent to the responses of any other respondent, these sums over $\mathbf{a}$ and $\mathbf{b}$ can be expanded as
\begin{equation}\label{eq:pryv2}
    \sum_{\mathbf{a}}\sum_{\mathbf{b}}=\sum_{a_1=0}^1\dots\sum_{a_s=0}^1\sum_{b_1=0}^1\dots\sum_{b_s=0}^1.
\end{equation}
Therefore, we can expand Eq.~\ref{eq:pryv1} further into
\begin{equation}
    {\rm Pr}(y)=\sum_{a_1=0}^1\dots\sum_{a_s=0}^1\delta_{y,\sum_i^s a_i}\sum_{b_1=0}^1\dots\sum_{b_s=0}^1{\rm Pr}(\mathbf{a},\mathbf{b}).
\end{equation}
Probability ${\rm Pr}(\mathbf{a},\mathbf{b})$ is fully factorized over each $i$, and therefore, we can begin to simplify it. Thus, we can first perform the sum of factor $i$ over $b_i$ as
\begin{equation}
    \sum_{b_i=0}^1\left[\kappa^{a_i}(1-\kappa)^{1-a_i}\phi\delta_{b_i,1}+\delta_{a_i,0}(1-\phi)\delta_{b_i,0}\right]
    =\delta_{a_i,0}(1-\phi)+\kappa^{a_i}(1-\kappa)^{1-a_i}\phi={\rm Pr}(a_i).
\end{equation}
This provides the expression
\begin{equation}\label{eq:pryv3}
    {\rm Pr}(y)=\sum_{a_1=0}^1\dots\sum_{a_s=0}^1\delta_{y,\sum_i^s a_i}\prod_{i=1}^s\left[\delta_{a_i,0}(1-\phi)+\kappa^{a_i}(1-\kappa)^{1-a_i}\phi\right]=\sum_{a_1=0}^1\dots\sum_{a_s=0}^1\delta_{y,\sum_i^s a_i}\prod_{i=1}^s{\rm Pr}(a_i).
\end{equation}
To complete this calculation, we now make the following observations. Since there are $s$ separate variables $a_1,\dots,a_s$ capable of contributing to produce $\sum_i a_i=y$, and each has two possible values, $0$ or $1$, obtaining $y$ from $\sum_i a_i$ is equivalent to picking $y$ among the $s$ variables to take on the value $1$ and the remaining $s-y$ to take the value $0$. This can be done in $\binom{s}{y}$ different ways, and any such combination makes the factor $\delta_{y,\sum_i a_i}$ become $1$. In other words, the summations $\sum_{\mathbf{a}}$ over all possible configurations of $\mathbf{a}$ lead to $\binom{s}{y}$ terms that can be non-zero on the basis of the delta on $y$. Furthermore, for any $a_i$ that takes on the value $1$, the corresponding factor in $\prod_i$ takes the value $\kappa\phi$; if $a_i=0$, the value taken by the factor is $1-\phi+(1-\kappa)\phi=1-\kappa\phi$. Since $y$ different $a_i$ are equal to $1$ and $s-y$ are $a_i=0$, these observations readily lead to
\begin{equation}\label{eq:pryv4}
    {\rm Pr}(y)=\binom{s}{y}(\kappa\phi)^y(1-\kappa\phi)^{s-y},
\end{equation}
a binomial distribution with success rate $\kappa\phi$, and expectation $\langle y\rangle=s\kappa\phi$. This equation corresponds to Eq.~4 of the main text.

The value of $\phi$ can be estimated from the survey.  This is done through a similar approach as above, by determining ${\rm Pr}(z)$, the probability for $z=\sum_i^s b_i$. It can be calculated in the same way as ${\rm Pr}(y)$, resulting in the binomial distribution
\begin{equation}\label{eq:prznossigma}
    {\rm Pr}(z)=\binom{s}{z}\phi^z(1-\phi)^{s-z}
\end{equation}
with success rate $\phi$ and expectation $\langle z\rangle=s\phi$.

\subsubsection{Derivations of location-level averages}\label{sec:model-location-averages}
We are now in a position to derive quantities that directly relate to the survey data used in our study. At this point, it is important to recall that both $\phi$ and $\kappa$ above are dependent on $g$, $\mathbf{c}$, and $\lambda$ even though this functional dependence has been abstracted away for notational simplicity. However, we know from our experience working with survey data that $s(g,\mathbf{c})$ can be small especially for less-populated $g$ if the sample is stratified. This means that $\kappa(g,\mathbf{c},\alpha)$ and $\phi(g,\mathbf{c})$ where both $g$ and $\mathbf{c}$ are specified may not be reliable for some locations due to small sample size. The situation would likely be worse if $\alpha$ is simultaneously specified. Thus, we must work at a greater level of aggregation, such as at the level $g$ where values of $\mathbf{c}$ are averaged over.

For location $g$ and activity-day $\alpha$, the model can be used to calculate the probability to obtain the collection $\mathbf{y}=(y_{\mathbf{c}_1},\dots,y_{\mathbf{c}_q})$ of counts of yes answers to the question of interaction with nclf. By the independence of the different $\mathbf{c}$, the joint probability of these answers is
\begin{equation}
    {\rm Pr}(\mathbf{y})=\prod_{\mathbf{c}}{\rm Pr}(y_{\mathbf{c}}),
\end{equation}
where we write the product over $\mathbf{c}$ instead of over the indices from $1$ to $q$ (simply to avoid adding more notation). The individual ${\rm Pr}(y_{\mathbf{c}})$ satisfy Eq.~\ref{eq:pryv3}, where it is understood that $\kappa=\kappa(g,\mathbf{c},\alpha)$ and $\phi(g,\mathbf{c})$ with, in general, different values for different $\mathbf{c}$. Due to independence, the expectation for this joint distribution is simply $\langle\mathbf{y}\rangle=\prod_{\mathbf{c}}\langle y_{\mathbf{c}}\rangle$. 

The results from the joint distribution can be used to determine a location-$g$ estimate of the fraction of people in the target population interacting with nclf. This estimate is proportional to the expectation of the random variable $\sum_{\mathbf{c}}w(g,\mathbf{c})y_{\mathbf{c}}$ which, by linearity and the independence of the random variable, is equal to $\sum_{\mathbf{c}}w(g,\mathbf{c})\langle y_{\mathbf{c}}\rangle=\sum_{\mathbf{c}}Q(g,\mathbf{c})\kappa(g,\mathbf{c},\alpha)\phi(g,\mathbf{c})$ after applying Eq.~\ref{eq:P_ws}. We have seen the quantity $\kappa(g,\mathbf{c},\alpha)\phi(g,\mathbf{c})$ above, and it represents the marginal likelihood that a respondent reports having interacted with nclf (the success rate in the binomial distribution in Eq.~\ref{eq:pryv4}). To make this last sum an expectation, we must divide by the total population $\sum_{\mathbf{c}}Q(g,\mathbf{c})=Q(g)$, which yields the expected fraction of the target population $f(g,\alpha)$ that reports performing activity-day $\alpha$ with nclf, or
\begin{equation}\label{eq:f_g_alpha_mod}
    f(g,\alpha)=\frac{\sum_{\mathbf{c}}Q(g,\mathbf{c})\kappa(g,\mathbf{c},\alpha)\phi(g,\mathbf{c})}{\sum_{\mathbf{c}}Q(g,\mathbf{c})}=\frac{\sum_{\mathbf{c}}Q(g,\mathbf{c})\kappa(g,\mathbf{c},\alpha)\phi(g,\mathbf{c})}{Q(g)}.
\end{equation}
This corresponds to an average interaction probability with nclf for activity-day $\alpha$ over individuals belonging to the target population in location $g$. Equation~\ref{eq:f_g_alpha_mod} is Eq.~5 of the main text, where we distinguish modeled and measured quantities by changing $f(g, \alpha)$ to $f_m(g, \alpha)$.

A similar analysis can be conducted on the answers to the question of nclf availability, leading to
\begin{equation}
    \phi(g)=\frac{\sum_{\mathbf{c}}Q(g,\mathbf{c})\phi(g,\mathbf{c})}{Q(g)},
\end{equation}
which corresponds to the average availability over individuals located in $g$. This is Eq.~6 of the main text.

The location-level $f(g,\alpha)$ and $\phi(g)$ can be combined to produce another useful quantity denoted by $\lambda(g)$ and defined as
\begin{equation}
    \lambda(g,\alpha)=\frac{f(g,\alpha)}{\phi(g)}=\frac{\sum_{\mathbf{c}}\kappa(g,\mathbf{c},\alpha)\phi(g,\mathbf{c})Q(g,\mathbf{c})}{\sum_{\mathbf{c}}\phi(g,\mathbf{c})Q(g,\mathbf{c})},
\end{equation}
which is Eq.~7 of the main text. 

There are three reasons for introducing $\lambda$. First, it can be directly estimated from data because the sample versions of $f$ and $\phi$ (given in Eqs.~\ref{eq:fhat} and~\ref{eq:phihat} below) depend directly on the data. Second, it provides an approximation to the average $\kappa$ per individual in $g$ that turns out to be the only such average that is conceptually sound. To be strict to the mathematics, the average contained in $\lambda(g,\mathbf{c})$ is not over all the target population in $g$, but rather the fraction of that population that has nclf available. This becomes clear from the fact that $\phi(g,\mathbf{c})Q(g,\mathbf{c})$ is the expectation of the sub-population $Q(g,\mathbf{c})$ who respond that they have nclf. The strict average of the $\kappa$ over $g$ and $\alpha$ is given by
\begin{equation}
    \kappa(g,\alpha)=\frac{\sum_{\mathbf{c}}\kappa(g,\mathbf{c},\alpha)Q(g,\mathbf{c})}{\sum_{\mathbf{c}}Q(g,\mathbf{c})}\neq\lambda(g,\alpha),
\end{equation}
where the non-equality is stated simply for clarity. The problem with $\kappa(g,\alpha)$ is twofold. From a practical standpoint we cannot directly calculate it accurately because our ability to estimate each $\kappa(g,\mathbf{c},\alpha)$ is very limited. More importantly, at the conceptual level, not everybody in $Q(g,\mathbf{c})$ actually has nclf with which to interact. Thus, it is not clear that $\kappa(g,\alpha)$ is defined properly in practice. Because of these conceptual considerations, we can think about $\lambda(g,\alpha)$ as an \textit{effective propensity}. The third reason to introduce $\lambda(g,\alpha)$ is because it provides a way to assess the trends in propensity as a function of log-population $p(g)$.

\subsection{Estimation using ATUS and PSTS data}\label{sec:estimators}
The results above, based on a stochastic model of respondents, combined with data from the ATUS and PSTS, can be used to make estimates of the parameters we are interested in.

Let us start by consider the most basic estimations from the model. Concentrating on a specific location $g$, activity-day $\alpha$, and population segment $\mathbf{c}$, we can formally write an estimation for $\kappa\phi$. By maximum likelihood estimation (MLE), if $y=\hat{y}$ is observed for this ($g,\mathbf{c},\alpha$)-combination and we want to determine the value of $\kappa\phi$ most likely to match that $y$, we take a derivative of ${\rm Pr}(y)$ with respect to the combined variable $\kappa\phi$ and seek its maximum. As usual, it is more convenient to do this with the logarithm of ${\rm Pr}(y)$, which leads to
\begin{equation}\label{eq:kappaphimle}
    \left.\frac{d}{d(\kappa\phi)}\log{\rm Pr}(y)\right|_{y=\hat{y}}=\frac{\hat{y}}{\hat{(\kappa\phi)}}-\frac{s-\hat{y}}{1-\hat{(\kappa\phi)}}=0\quad\Longrightarrow\quad\hat{(\kappa\phi)}=\frac{\hat{y}}{s}=\frac{\sum_{i=1}^s\hat{a}_i}{s},
\end{equation}
where $\hat{a}_i$ are the respondent answers, converted to $0$ or $1$, recorded in the ATUS about interaction with nclf in the group of respondents with $(g,\mathbf{c},\alpha)$. In other words, ${\rm Pr}(y=\hat{y})$ is maximized if $\kappa\phi$ is given by Eq.~\ref{eq:kappaphimle}.

Taking the MLE approach, we can estimate $\phi$ on the basis of the PSTS family availability data (see SI~\ref{sec:si-availability}) on $\hat{z}=\sum_i^{s}\hat{b}_i$ and the number of respondents in each $g$ with $\mathbf{c}$ to find
\begin{equation}
    \left.\frac{d}{d\phi}\log{\rm Pr}(z)\right|_{z=\hat{z}}=\frac{\hat{z}}{\hat{\phi}}-\frac{s-\hat{z}}{1-\hat{\phi}}=0\quad\Longrightarrow\quad\hat{\phi}=\frac{\hat{z}}{s}=\frac{\sum_{i=1}^s \hat{b}_i}{s}.
\end{equation}

As discussed in Sec.~\ref{sec:model-location-averages}, we cannot actually take full advantage of these estimates because of small sample sizes for many locations if the samples are stratified into characteristics $\mathbf{c}$. Nevertheless, we can use them to empirically estimate $f(g,\alpha)$, achieved by inserting Eq.~\ref{eq:kappaphimle} into Eq.~\ref{eq:f_g_alpha_mod} while taking into account Eq.~\ref{eq:P_ws}, producing
\begin{equation}\label{eq:fhat}
    \hat{f}(g,\alpha)=\frac{\sum_{\mathbf{c}}w(g,\mathbf{c})\sum_{i}^{s(g,\mathbf{c})}\hat{a}_i(\alpha)}{P(g)}=\frac{\sum_{\mathbf{c}}w(g,\mathbf{c})\sum_{i}^{s(g,\mathbf{c})}\hat{a}_i(\alpha)}{\sum_{\mathbf{c}}w(g,\mathbf{c})s(g,\mathbf{c})},
\end{equation}
where we explicitly highlight that the collected answer for $\hat{a}_i$ is in reference to activity-day $\alpha$. 
Similarly,
\begin{equation}\label{eq:phihat}
    \hat{\phi}(g)=\frac{\sum_{\mathbf{c}}w(g,\mathbf{c})\sum_i^{s(g,\mathbf{c})}\hat{b}_i}{\sum_{\mathbf{c}}w(g,\mathbf{c})s(g,\mathbf{c})}.
\end{equation}

The Eqs. for $\hat{f}(g,\alpha)$ and $\hat{\phi}(g)$ can be used directly to calculate population-weighted averages over the entire set of locations for interaction due to a specific activity-day (sample result labeled $\hat{f}(\alpha)$) and availability (sample result labeled $\hat{\phi}$). They also make it possible to provide the estimator $\hat{\lambda}(g,\alpha)$. However, to write it explicitly, we must remember that $\hat{\phi}(g)$ is derived from the Pew data whereas $\hat{f}(g,\alpha)$ is obtained from the ATUS. Thus, we estimate with
\begin{equation}
    \hat{\lambda}(g,\alpha)=\frac{\hat{f}(g,\alpha)}{\hat{\phi}(g)}.\quad\begin{array}{l}
         \leftarrow \text{from ATUS}\\
         \leftarrow \text{from Pew}
    \end{array}
\end{equation}

As a general statement, when the variable $a_i$ is involved, the ATUS sample sizes and weights need to be used. When $b_i$ is involved, the Pew survey sample sizes and weights are needed.

\section{Estimation of population-size dependence of interaction, availability, and propensity through non-parametric modal regression with KDE} \label{sec:estimation}
One of our approaches to understand how $f$, $\phi$, $\lambda$, and $t$ are affected by population is non-parametric modal regression~\cite{ChenMode}. To describe the technique in detail, consider a set of random variables $\{u_g,v_g\}_g$ where $u_g$ is considered the independent variable. The goal of the method is to find a function $v$ of $u$ that represents the relation between $\{u_g,v_g\}_g$.

The method first estimates the bi-variate probability density $\rho(u,v)$ using kernel density estimation. In our case, we employ a weighted Gaussian kernel estimation where the selection of the bandwidth is explained below. The algorithm we use is \texttt{gaussian\_kde}, part of the \texttt{python} package \texttt{scipy.stats}, and the weights of the data points are those calculated for each $g$ as explained in the main text, Methods section. In our implementation, the $u,v$ space is discretized into $n_u\times n_v$ tiles, each with a value $\rho_{i,j}$ corresponding to the density $\rho(u_i,v_j)$ where $u_i$ is a value of $u$ equal to $u_{\min}+i\times du$ and $i$ is an integer between $1$ and $n_u$, and $v_j$ is a value of $v$ equal to $v_{\min}+j\times dv$ and $j$ is an integer between $1$ and $n_v$. The values $u_{\min}$ and $v_{\min}$ are, respectively the minimum values of $u$ and $v$ in the region of interest. The two parameters $du$ and $dv$ are estimated using $du=(u_{\max}-u_{\min})/n_u$ and $dv=(v_{\max}-v_{\min})/n_v$ where $u_{\max}$ and $v_{\max}$ are, respectively the maximum values of $u$ and $v$ in the region of interest. 

After determining the density, we construct the conditional density $\rho(v| u)=\rho(u,v)/\rho(u)$ where $\rho(u)$ is the marginal density for variable $u$. This marginal is numerically estimated from the joint $\rho(u,v)$ by adding its values for fixed $u$. In our discretization, $\rho(u)$ is calculated by performing the sum $\sum_{j=1}^{n_v}\rho_{i,j}$ to yield $\rho_{i}$ which corresponds to $\rho(u_i)$. This is then used in $\rho_{i,j}/\rho_{i}$ which corresponds to $\rho(v_j|u_i)$. 

To construct the modal regression, for a given $u_i$ we locate the $j$, labeled $j^*$, that leads to the largest value of the set $\{\rho(v_j|u_i)\}_{j}$. This $j^*$ is a function of $i$, that is $j^*(i)$. Therefore, the modal regression corresponds to the set of values $\{u_i,j^*(u_i)\}_i$ which, in terms of $u$ and $v$, takes the form $\{u_i,v_{j^*(i)}\}_i$ or in a more familiar way, $(u_i,v^*(u_i))$. Figure~1 of the main text, and all the versions of the same figure presented here in the SI as robustness checks are constructed with this method (excluding the smoothing splines, explained in the Methods of the main text), where the color corresponds to $\rho(v_j|u_i)$ and the modal regression to $(u_i,v^*(u_i))$. In all cases, $u$ corresponds to the value of log-population $p(g)=\log P(g)$. On the other hand, $v$ takes on four different forms, $f$, $\phi$, $\lambda$, and $t$. The values for $n_u$ and $n_v$ are both equal to $2,000$. 

\begin{figure}
    \centering
    \includegraphics[scale=0.63]{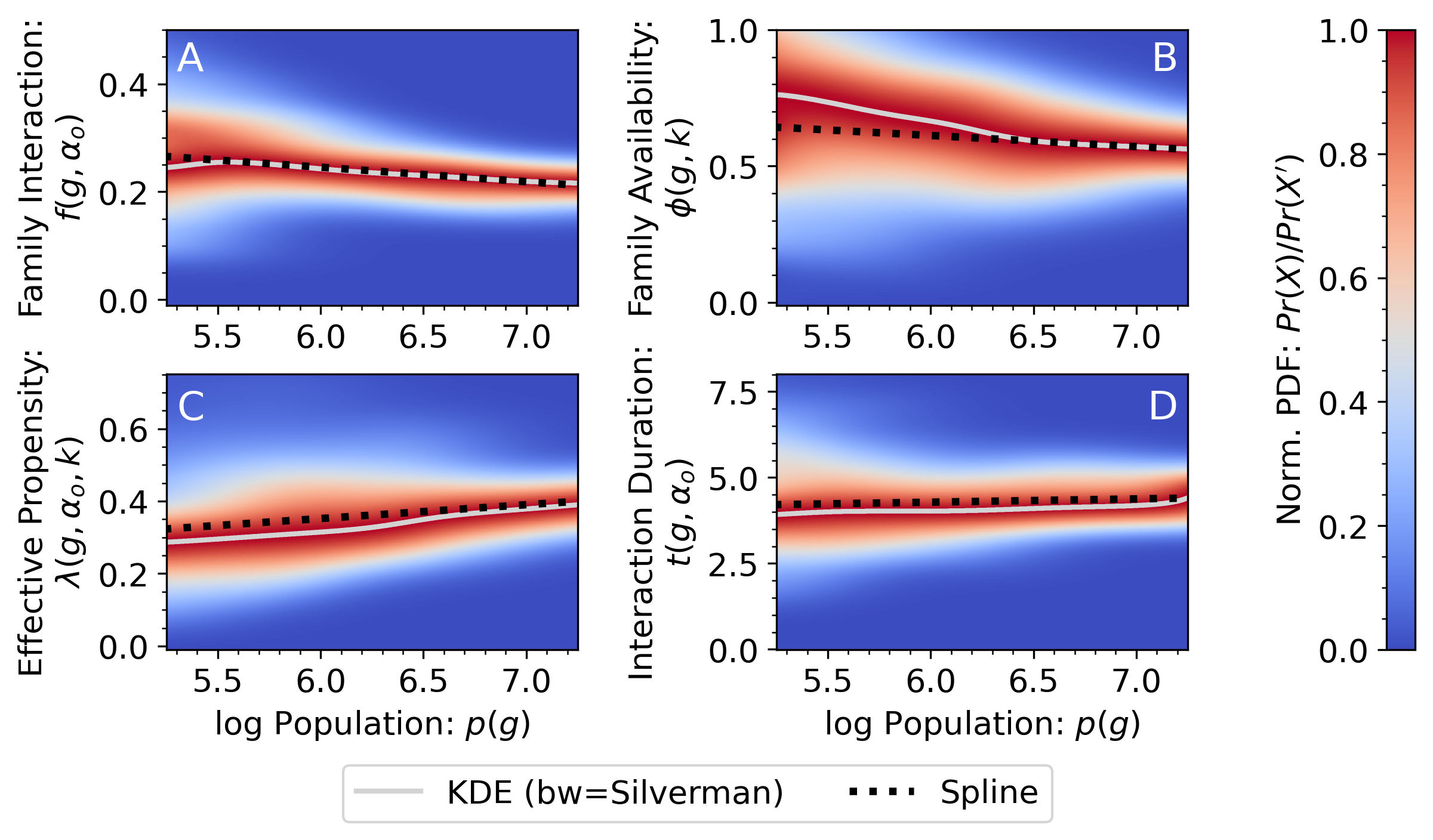}
    \includegraphics[scale=0.63]{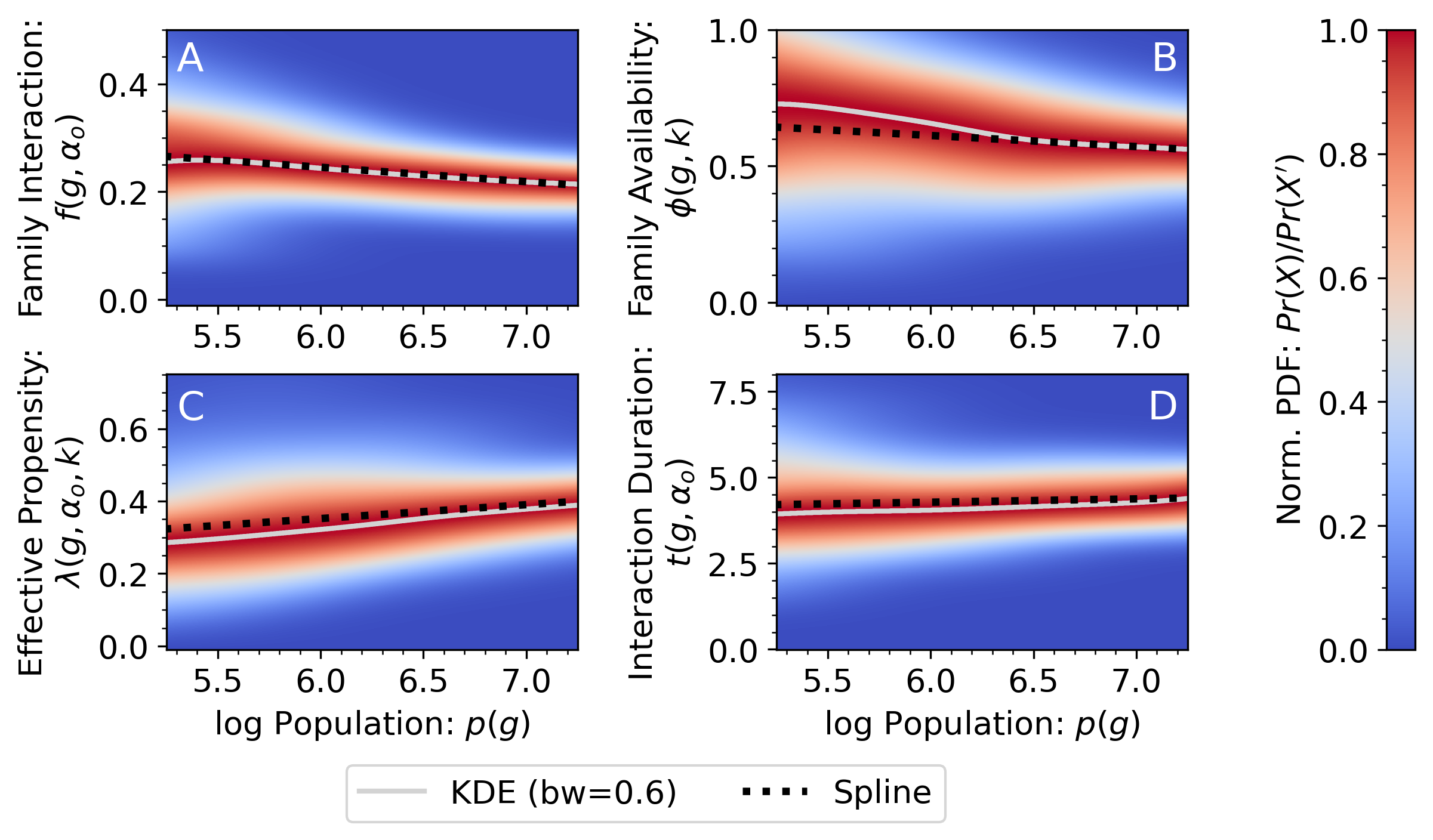}
    \includegraphics[scale=0.63]{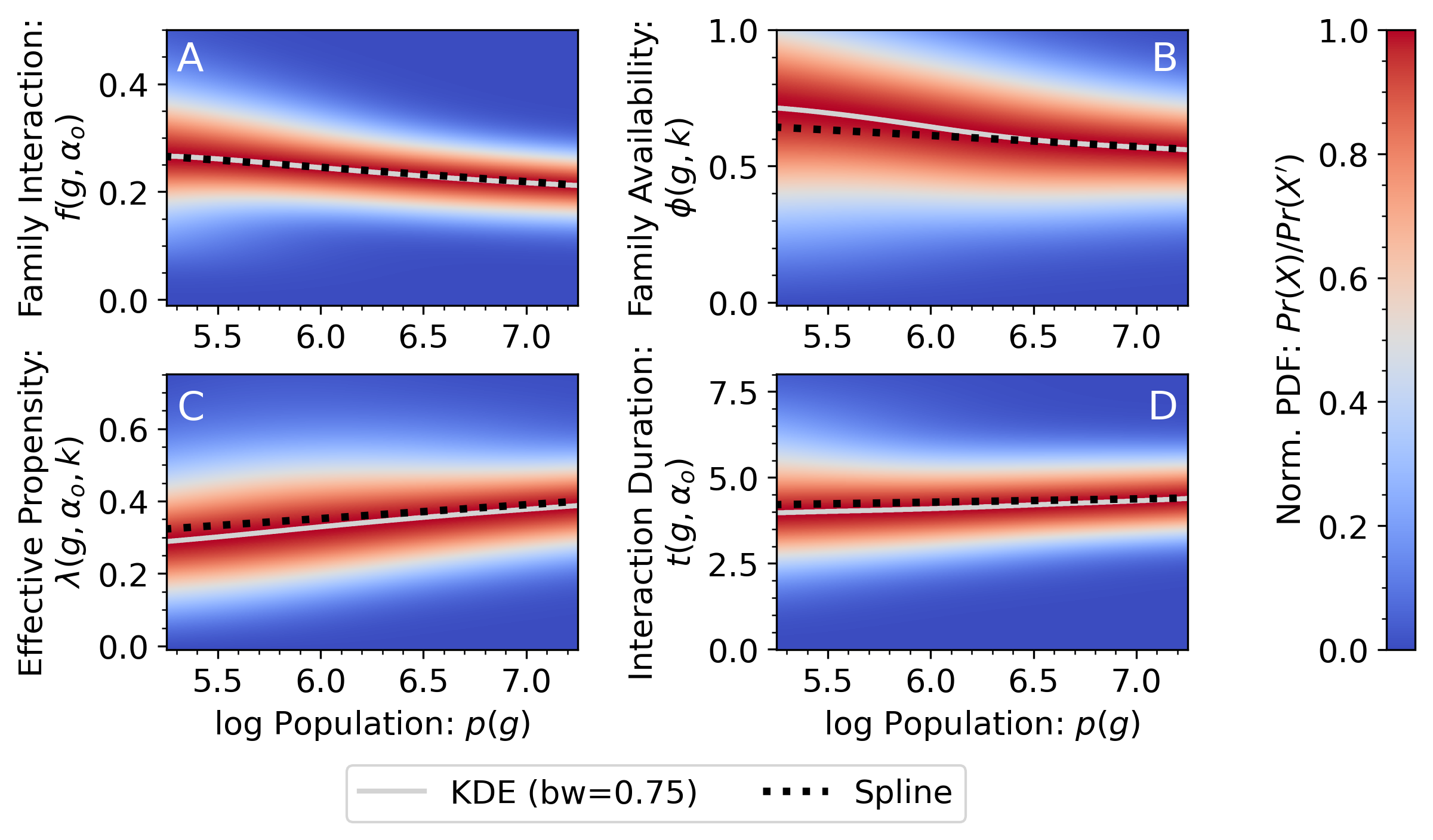}
    \caption{Versions of Fig.~1 of the main text with different choices of bandwidth. Top corresponds to Silverman's bandwidth $0.497$, middle to $0.6$, and bottom to $0.75$, our chosen value. The larger bandwidth leads to the modal regression changing slightly for small $p$, leading to more similarity with the results of the smoothing splines method.}
    \label{fig:bandwidth}
\end{figure}
Bandwidth selection for the KDE is estimated by inspection in the neighborhood of the Silverman bandwidth~\cite{silverman1986density}, of value $\approx 0.497$. The choice is made to enhance stability of the model regression, and we settle on a value of $0.75$. Fig.~\ref{fig:bandwidth} shows how the modal regression varies as a consequence of varying the bandwidth. The effect of larger bandwidth does not change any of the qualitative results. The modal regression becomes more similar to the smooth spline which is consistent with our intention to increase stability.

\section{Summary Tables}
\subsection{Average propensity per activity}
It is of interest to know which specific activities people tend to be more inclined towards. To examine this directly, we average each $\lambda$ for each $\alpha$ over the cities, weighted by sample size, to provide an expected value for each activity. We show the results, sorted by value, in Tab.~\ref{tab:avg_lambda}, with aggregate $\alpha$s shown in bold. Concretely, we compute
\begin{equation}
    \langle\lambda(\alpha,k)\rangle=\frac{\sum_g w_g \lambda(g,\alpha,k)}{\sum_{g} w_g}
\end{equation}
where $w_g$ is the city sample size, explained in the Methods section, main text.

To categorically describe these results, it appears that social activities drive the majority of propensity to interact. These are followed by care and home-centered activities. Following these, we observe a mixture of activities that are either rare themselves, or common but either typically conducted alone, or where individuals have less control over who they conduct the activity alongside. Among these we see activities such as education and work, as well as personal care (hygiene) and civic activities (voting, seeking government assistance).

\begin{table}
\centering
\begin{tabular}{llrrr}
\toprule
                    Activity &  ATUS Codes &  &Avg Propensity & \\\cmidrule{3-5}%
                   $\alpha$ &    &  $\langle\lambda(\alpha, k=1)\rangle$ & $\langle\lambda(\alpha, k=2)\rangle$ &  $\langle\lambda(\alpha, k=3)\rangle$ \\
\midrule
            \textbf{Any} &  All codes &     0.2894 &   0.3307 &   0.3639 \\
         \textbf{Social} & 11, 12, 13 &     0.2458 &   0.2809 &   0.3093 \\
       Social \& Leisure &     12     &     0.1978 &   0.2265 &   0.2500 \\
      Eating \& Drinking &     11     &     0.1504 &   0.1717 &   0.1891 \\
               Traveling &     18     &     0.1119 &   0.1279 &   0.1411 \\
           \textbf{Care} & 03, 04     &     0.0868 &   0.0997 &   0.1105 \\
              Household  &     02     &     0.0681 &   0.0781 &   0.0865 \\
       Care, Non-Coresid &     04     &     0.0661 &   0.0759 &   0.0841 \\
      Consumer Purchases &     07     &     0.0418 &   0.0479 &   0.0531 \\
           Care, Coresid &     03     &     0.0247 &   0.0285 &   0.0317 \\
     Sports, Exrc \& Rec &     13     &     0.0200 &   0.0228 &   0.0252 \\
               Religious &     14     &     0.0148 &   0.0172 &   0.0193 \\
                    Work &     05     &     0.0085 &   0.0097 &   0.0107 \\
 Prof. \& Pers. Services &     08     &     0.0061 &   0.0070 &   0.0077 \\
             Phone calls &     16     &     0.0048 &   0.0055 &   0.0062 \\
               Volunteer &     15     &     0.0046 &   0.0053 &   0.0059 \\
               Education &     06     &     0.0026 &   0.0031 &   0.0035 \\
         Household Serv. &     09     &     0.0019 &   0.0022 &   0.0025 \\
           Personal Care &     01     &     0.0017 &   0.0019 &   0.0022 \\
     Govt Serv. \& Civic &     10     &     0.0005 &   0.0006 &   0.0006 \\
\bottomrule
\end{tabular}
\caption{Rank-ordered list of propensity values by activity over all major activity categories captured in the ATUS, for values of $k$ ranging from 1 to 3. The numerical codes are those stated in the ATUS Lexicon of 2019~\cite{ATUSlex}. Aggregate activities are highlighted in bold and the codes used to construct them are indicated. }
\label{tab:avg_lambda}
\end{table}

\subsection{Listing MSAs in terms of $\lambda$ and $t$}
In order to measure how much a city $g$ with $p(g)$ differs from the typical behavior of other cities with similar $p\approx p(g)$, we introduce a weighted $z$-score that adjusts for the heteroskedasticity of our data. The specific definition is as follows. From the estimated bivariate conditional density $\rho(v|u)$ of the set of values $\{u_g,v_g\}_g$, determined as explained above in Sec.~\ref{sec:estimation}, we calculate the expectation
\begin{equation}
    \langle v(u_g)\rangle=\langle v(u_{i(g)})\rangle=\frac{\sum_{j=1}^{n_v}v_j\times \rho(v_j|u_{i(g)})}{n_v},
\end{equation}
where $u_{i(g)}$ is $u_i$ with $i$ that satisfies $u_{i}\leq u_g<u_{i+1}$. Similarly, the standard deviation is estimated from the usual sample formula
\begin{equation}
    \sigma_{v(u_g)}=\sqrt{\frac{\sum_{j=1}^{n_v}[v_j-\langle v(u_g)\rangle]^2}{n_v}}.
\end{equation}
Finally, in order to determine the $z$-score, we apply
\begin{equation}\label{eq:w-z}
    z(v(u_g))=\frac{w_g}{\sum_g w_g}\times\frac{v(u_{i(g)})-v^*(u_{g})}{\sigma_{v(u_g)}}
\end{equation}

To apply this result, $u$ always corresponds to $p$ and $v$ is either the effective propensity $\lambda$ or the interaction duration $t$. 
The inclusion of weights here further allows us to emphasize cities that are well-sampled, for which we show the top and bottom ranked cities in terms of $\lambda(g, \alpha_o, k=3)$ in Tab.~\ref{tab:lam}, and the same selection criteria for $t(g, \alpha_o)$ in Tab.~\ref{tab:t}.

\begin{table}[t]
\centering
\begin{tabular}{rllll}
\toprule
 CBSA &                                         Name & \multicolumn{3}{c}{$\lambda(g,\alpha_o,k)$ (rank by weighted $z$-score)}  \\\cmidrule{3-5}%
      &                                              & $k=1$ & $k=2$ & $k=3$ \\
\midrule
38060 &                  Phoenix-Mesa-Scottsdale, AZ &        0.389 (1) &        0.473 (1) &        0.563 (1) \\
35620 &        New York-Newark-Jersey City, NY-NJ-PA &        0.297 (4) &        0.350 (2) &        0.403 (2) \\
33460 &      Minneapolis-St. Paul-Bloomington, MN-WI &        0.336 (3) &        0.386 (3) &        0.432 (3) \\
16740 &            Charlotte-Concord-Gastonia, NC-SC &        0.396 (2) &        0.427 (4) &        0.453 (4) \\
19740 &                   Denver-Aurora-Lakewood, CO &        0.422 (5) &        0.506 (5) &        0.592 (5) \\
45300 &          Tampa-St. Petersburg-Clearwater, FL &        0.363 (6) &        0.444 (7) &        0.530 (8) \\
26900 &             Indianapolis-Carmel-Anderson, IN &        0.329 (9) &        0.428 (6) &        0.548 (6) \\
36420 &                            Oklahoma City, OK &        0.436 (8) &        0.559 (8) &       0.704 (10) \\
33100 &    Miami-Fort Lauderdale-West Palm Beach, FL &        0.330 (7) &        0.384 (9) &       0.437 (11) \\
41180 &                             St. Louis, MO-IL &       0.309 (14) &       0.367 (10) &        0.426 (9) \\
 \multicolumn{1}{c}{\vdots}&  &  \multicolumn{1}{c}{\vdots}&  \multicolumn{1}{c}{\vdots} &  \multicolumn{1}{c}{\vdots} \\
44700 &                            Stockton-Lodi, CA &      0.195 (200) &      0.195 (204) &      0.195 (207) \\
21660 &                                   Eugene, OR &      0.156 (199) &      0.156 (207) &      0.156 (208) \\
46060 &                                   Tucson, AZ &      0.178 (211) &      0.221 (208) &      0.267 (195) \\
16700 &              Charleston-North Charleston, SC &      0.199 (207) &      0.214 (206) &      0.227 (206) \\
49340 &                             Worcester, MA-CT &      0.204 (208) &      0.204 (210) &      0.204 (212) \\
40900 &      Sacramento--Roseville--Arden-Arcade, CA &      0.241 (209) &      0.255 (211) &      0.265 (211) \\
28140 &                           Kansas City, MO-KS &      0.227 (210) &      0.259 (212) &      0.289 (209) \\
47900 & Washington-Arlington-Alexandria, DC-VA-MD-WV &      0.247 (214) &      0.298 (213) &      0.351 (205) \\
31080 &           Los Angeles-Long Beach-Anaheim, CA &      0.265 (212) &      0.314 (214) &      0.364 (213) \\
14460 &               Boston-Cambridge-Newton, MA-NH &      0.200 (215) &      0.238 (215) &      0.276 (215) \\
\bottomrule
\end{tabular}
\caption{Average Rank-ordered listings of the top and bottom ten CBSAs in the US for $\lambda$ on the basis of weight-adjusted $z$-scores as defined by Eq.~\ref{eq:w-z}, for values of $k$ ranging from 1 to 3. }
\label{tab:lam}
\end{table}

\begin{table}[t]
\centering
\begin{tabular}{rlrr}
\toprule
 CBSA &                                         Name &      $t(g,\alpha_o)$ &  Rank by \\
    &                                            &   &  weighted $z$-score \\
\midrule
  47900 & Washington-Arlington-Alexandria, DC-VA-MD-WV & 5.2334 &       1 \\
  35620 &        New York-Newark-Jersey City, NY-NJ-PA & 4.7812 &       2 \\
  42660 &                  Seattle-Tacoma-Bellevue, WA & 5.6429 &       3 \\
  26420 &         Houston-The Woodlands-Sugar Land, TX & 4.8333 &       4 \\
  38060 &                  Phoenix-Mesa-Scottsdale, AZ & 4.9774 &       5 \\
  19820 &                  Detroit-Warren-Dearborn, MI & 4.9197 &       6 \\
  41700 &                San Antonio-New Braunfels, TX & 5.3979 &       7 \\
  41860 &            San Francisco-Oakland-Hayward, CA & 4.9484 &       8 \\
  46140 &                                    Tulsa, OK & 6.1930 &       9 \\
  26900 &             Indianapolis-Carmel-Anderson, IN & 4.7501 &      10 \\
 \multicolumn{1}{c}{\vdots} &                                              &  \multicolumn{1}{c}{\vdots} &  \vdots \hspace{2mm} \\
  19100 &              Dallas-Fort Worth-Arlington, TX & 4.0465 &     205 \\
  39580 &                                  Raleigh, NC & 3.1891 &     206 \\
  36740 &                Orlando-Kissimmee-Sanford, FL & 3.4537 &     207 \\
  45300 &          Tampa-St. Petersburg-Clearwater, FL & 3.6301 &     208 \\
  41620 &                           Salt Lake City, UT & 3.1777 &     209 \\
  41740 &                       San Diego-Carlsbad, CA & 3.5644 &     210 \\
  16980 &           Chicago-Naperville-Elgin, IL-IN-WI & 4.0296 &     211 \\
  36420 &                            Oklahoma City, OK & 3.0496 &     212 \\
  31080 &           Los Angeles-Long Beach-Anaheim, CA & 4.0002 &     213 \\
  17460 &                         Cleveland-Elyria, OH & 2.6093 &     214 \\
\bottomrule
\end{tabular}
\caption{Rank-ordered listings of the top and bottom ten CBSAs in the US for $t$ on the basis of weight-adjusted $z$-scores as defined by Eq.~\ref{eq:w-z}. }
\label{tab:t}
\end{table}

\clearpage
\bibliography{supp}